\documentclass[
 reprint,
 floatfix,
superscriptaddress,
 amsmath,amssymb,
 aps,prl
]{revtex4-2}
\usepackage{xcolor}
\usepackage{graphicx}
\usepackage{dcolumn}
\usepackage{bm}
\usepackage{subfig}
\usepackage{soul}
\usepackage{float}

\usepackage{ragged2e}
\DeclareCaptionJustification{justified}{\justifying}
\begin{document}
\raggedbottom
\preprint{APS/123-QED} 

\title{Collisions of Spin-polarized YO Molecules for Single Partial Waves}

\author{Justin J. Burau}
\affiliation{
JILA, National Institute of Standards and Technology and the University of Colorado, Boulder, Colorado 80309-0440 \\ 
Department of Physics, University of Colorado, Boulder, Colorado 80309-0390, USA
}

\author{Kameron Mehling}
\affiliation{
JILA, National Institute of Standards and Technology and the University of Colorado, Boulder, Colorado 80309-0440 \\ 
Department of Physics, University of Colorado, Boulder, Colorado 80309-0390, USA
}

\author{Matthew D. Frye}
\affiliation{Joint Quantum Centre (JQC), Durham - Newcastle, Department of Chemistry,
Durham University, Durham DH1 3LE, United Kingdom}
\affiliation{Faculty of Physics, University of Warsaw, Pasteura 5, 02-093 Warsaw, Poland}

\author{Mengjie Chen}
\affiliation{
JILA, National Institute of Standards and Technology and the University of Colorado, Boulder, Colorado 80309-0440 \\ 
Department of Physics, University of Colorado, Boulder, Colorado 80309-0390, USA
}

\author{Parul Aggarwal}
\affiliation{
JILA, National Institute of Standards and Technology and the University of Colorado, Boulder, Colorado 80309-0440 \\ 
Department of Physics, University of Colorado, Boulder, Colorado 80309-0390, USA
}

\author{Jeremy M. Hutson}
\affiliation{Joint Quantum Centre (JQC), Durham - Newcastle, Department of Chemistry,
Durham University, Durham DH1 3LE, United Kingdom}

\author{Jun Ye}
\affiliation{
JILA, National Institute of Standards and Technology and the University of Colorado, Boulder, Colorado 80309-0440 \\ 
Department of Physics, University of Colorado, Boulder, Colorado 80309-0390, USA
}

\date{\today}

\begin{abstract}
Efficient sub-Doppler laser cooling and optical trapping of YO molecules offer new opportunities to study collisional dynamics in the quantum regime. Confined in a crossed optical dipole trap, we achieve the highest phase-space density of $2.5 \times 10^{-5}$ for a bulk laser-cooled molecular sample. This sets the stage to study YO--YO collisions in the microkelvin temperature regime, and reveal state-dependent, single-partial-wave two-body collisional loss rates. We determine the partial-wave contributions to loss of specific rotational states (first excited $N=1$ and ground $N=0$) following two strategies. First, we measure the change of the collision rate in a spin mixture of $N=1$ by tuning the kinetic energy with respect to the p- and d-wave centrifugal barriers. Second, we compare loss rates between a spin mixture and a spin-polarized state in $N=0$. Using quantum defect theory with a partially absorbing boundary condition at short range, we show that the dependence on temperature for $N=1$ can be reproduced in the presence of a d-wave or f-wave resonance, and the dependence on a spin mixture for $N=0$ with a p-wave resonance.
\end{abstract}

\maketitle

There has been immense progress recently in the field of dipolar quantum gases of molecules, culminating in the achievement of both degenerate Fermi gases~\cite{Marco2019,Valtolina2020,Schindewolf2022, Pan2022} and Bose-Einstein condensate~\cite{bigagli2023observation}. The strong electric dipole moments of these dipolar gases have allowed the study of dipole-mediated spin many-body dynamics~\cite{Li2023} with a further promise to study strongly correlated phases \cite{Correlated,longrange} and exotic states of matter \cite{Triangle,dipolarlattice,bound}. However, these degenerate gases have thus far been limited to molecules associated from pairs of alkali-metal atoms. An alternative route, the direct laser cooling of molecules~\cite{eric,Truppe_2017,3dYO,Cafradio,Vilas_2022}, allows a complementary set of neutral molecules to be cooled to the ultracold regime~\cite{ding2020,caflambda,deep,magnetic} for a wide range of applications including the search for the electron's electric dipole moment~\cite{Roussy2023, ACME2018} and ultracold chemistry ~\cite{Liu_2021,Son_2022,Chen2024}.

\begin{figure}[t!]
\includegraphics[width= 1 \linewidth]{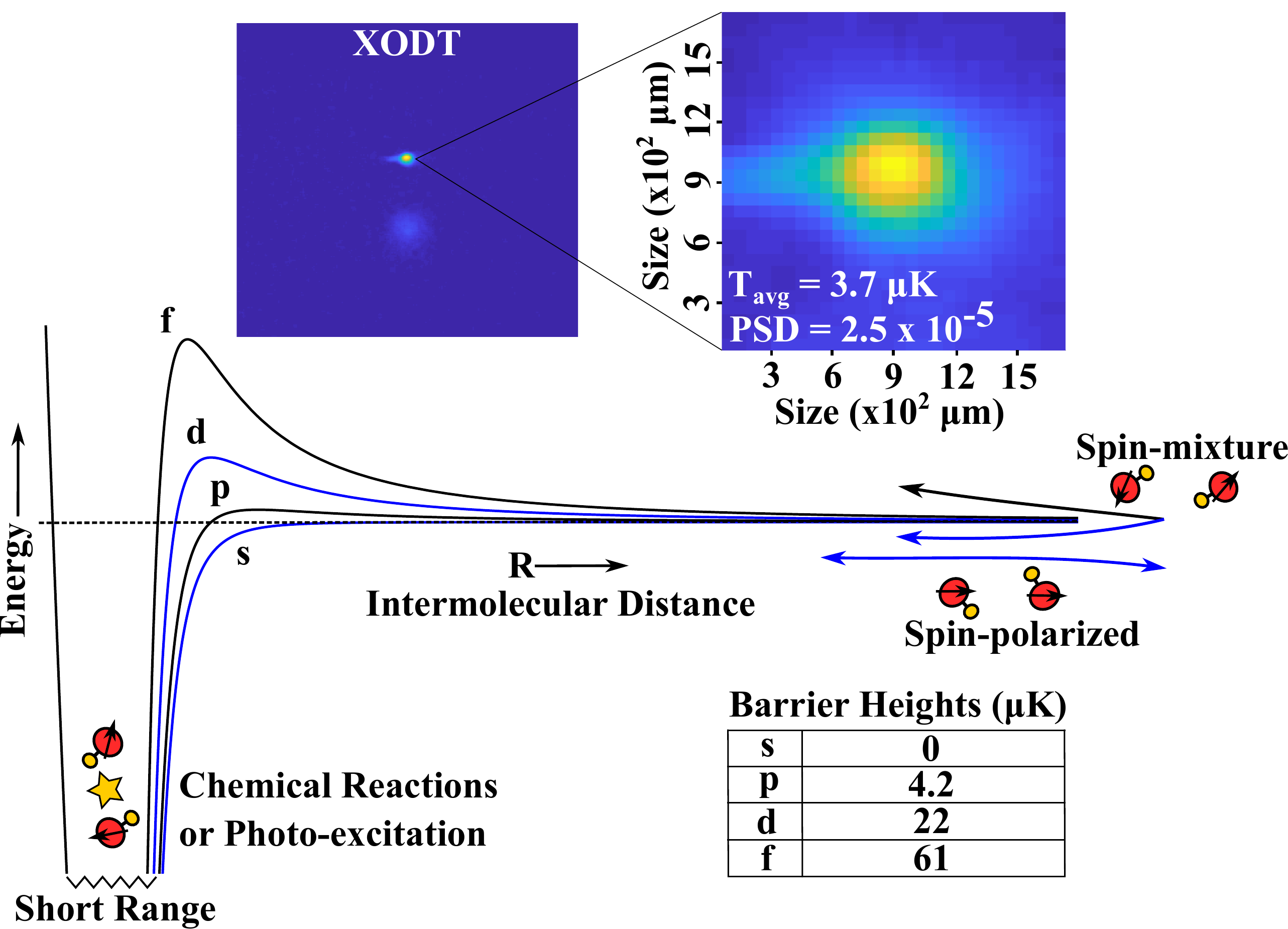}
\caption{
\label{fig:van der} Top: Images of YO molecules loaded into a crossed optical dipole trap (XODT). The unloaded molecules are seen falling, and a record phase-space density of $2.5 \times 10^{-5}$ is achieved inside the XODT. Bottom: Molecules interact through a long-range van der Waals potential. At sufficiently low collision energies, distinguishable molecules enter through both even (s,d) and odd (p,f) partial-wave channels, where partial waves beyond the s-wave feature a centrifugal barrier. The corresponding centrifugal barrier heights are shown in the table. Since YO is a boson, indistinguishable molecules collide only via even partial waves. Upon reaching short range, molecules undergo collisional loss. 
}
\end{figure}

The achievement of quantum degeneracy through evaporative cooling is facilitated by two ingredients: an initially high phase-space density and a good ratio of elastic to inelastic collisions ~\cite{Bec,ketterle}. Thus far, no directly laser-cooled molecules have been brought to quantum degeneracy, largely due to the initial phase-space density of the molecular gas being insufficiently high. A main limiting factor was low transfer efficiencies from a magneto-optical trap (MOT) into conservative traps \cite{Wu2021}, but this has since been resolved with the demonstration of the first sub-Doppler blue-detuned molecular MOT for YO molecules~\cite{Burau2023}. Subsequently, blue-detuned MOTs have also been demonstrated for SrF~\cite{jorapur2023high}, CaF~\cite{li2023bluedetuned}, and CaOH~\cite{hallas2024high} molecules, with bulk-gas collisions reported in an optical dipole trap (ODT) of SrF~\cite{jorapur2023high}.

Shielding of detrimental inelastic collisions has been proposed and demonstrated via dc electric~\cite{Wang_2015electric,electricBohn,Valtolina2020,Matsuda_2020,cafelectric} and microwave fields~\cite{tjis,Anderegg2021, Hongkong2023}, resulting in efficient evaporative cooling of bialkali dipolar molecules~\cite{Valtolina2020,Li2021,Schindewolf2022,Pan2022,bigagli2023observation}. With a large dipole moment of 4.5 D in the molecular frame, we expect YO to be amenable to both methods of shielding, ensuring prospects for evaporation.  

In this Letter, we report the loading of laser-cooled YO molecules into a crossed ODT (XODT), with a record phase-space density of $2.5 \times 10^{-5}$. This achievement is based on robust cooling of YO in both free space and within optical traps, due to its favorable ground-state level structure. YO features a large Fermi-contact interaction between the electron and nuclear spin, resulting in a separation of 760 MHz between the $G = 0$ and $G = 1$ hyperfine manifolds~\cite{ding2020}. This structure supports efficient gray molasses cooling (GMC) and we routinely achieve a temperature of 1.5 $\mu$K in free space, and a few $\mu$K in an XODT (as shown at the top of Fig.\ \ref{fig:van der}). This provides favorable conditions to explore YO collision dynamics by tuning the thermal energy to control different partial-wave contributions to the collisional loss rates. 
We model the rate coefficients using quantum defect theory (QDT), and show that the temperature dependence can be reproduced in the presence of a d-wave or f-wave scattering resonance. Further, we spin-polarize YO molecules in the absolute ground state $N = 0, G = 1, F = 1$ with the use of microwave sweeps between $N = 1$ and $N = 0$. We compare the collision rates for spin-polarized molecules and an incoherent spin mixture of hyperfine sublevels. The ratio of the rate coefficients for polarized and unpolarized samples can be reproduced in the presence of a p-wave resonance.

In the absence of external static or oscillating electric fields that polarize the heteronuclear molecules in the laboratory frame, the intermolecular interaction is well described by the long-range van der Waals potential
\begin{equation}
    V(R) = -\frac{C_6}{R^6} + \frac{\hbar^2L(L+1)}{2\mu R^2}
        \label{eq:indistingusihable},
\end{equation}
where $\mu$ is the reduced mass. The low-energy collision dynamics are dominated by the resulting centrifugal barriers for different partial waves $L$.
As depicted in the bottom of Fig.\ \ref{fig:van der}, the symmetrization requirement of the wave-function for a pair of bosonic YO molecules dictates that they interact through even partial waves when they occupy the same internal state. Distinguishable molecules can enter through both even and odd partial waves. For a partial wave that features a centrifugal barrier, the colliding molecules can reach short range only if they can tunnel through the barrier or have sufficient kinetic energy to overcome it. This results in a strong energy dependence on the scale of the barrier heights.

To illustrate the effect of these collision channels on total loss rate, we begin with the case of identical bosons subject only to two-body loss according to the form
\begin{equation}
    \dot{n} =  -(2 k_\textrm{e}) n^2 
        \label{eq:indistingusihablebegin},
\end{equation}
where $ \dot{n}$ represents the density loss rate, and $k_\textrm{e}$ is the even-partial-wave loss rate coefficient. The factor of 2 in Eq.\ (\ref{eq:indistingusihablebegin}) arises because each inelastic collision removes two molecules from the trap~\footnote{Note that some conventions \cite{chengchin} incorporate a factor of 2 within the definition of the cross-section (and so rate coefficient) but not in the rate equation. However, here we follow the convention of Huo and Green \cite{Huo:1996} which places this factor in the rate equation Eq.(\ref{eq:indistingusihablebegin}) and is usually used in molecular scattering as it is clearer for deriving relations with multiple occupied states, and for imposing correct symmetrization.}.

In contrast, molecular loss in a spin mixture is modeled by a set of coupled rate equations. For example, the total density for a mixture of three different spin states is
\begin{equation}
    n =  n_1 + n_2 +n_3
        \label{eq:total}.
\end{equation}
Collisions for both even and odd partial waves occur between different spin states while only even partial waves occur within individual spin states. If there are no purely spin-changing collisions, the two-body loss rate equations for each spin state are
\begin{equation}
    \begin{aligned}
    \dot{n}_1 =  -(2 k_\textrm{e}) n_1^2  - (k_\textrm{e} +k_\textrm{o})n_1n_2-(k_\textrm{e} +k_\textrm{o})n_1n_3 , \\
    \dot{n}_2 =  -(2 k_\textrm{e}) n_2^2  - (k_\textrm{e} +k_\textrm{o})n_2n_1-(k_\textrm{e} +k_\textrm{o})n_2n_3 , \\
    \dot{n}_3 =  -(2 k_\textrm{e}) n_3^2  - (k_\textrm{e} +k_\textrm{o})n_3n_2-(k_\textrm{e} +k_\textrm{o})n_3n_1,
    \end{aligned}
        \label{eq:distingusihable1}
\end{equation}
where we have introduced the loss rate coefficient $k_\textrm{o}$ for odd partial waves, and assumed both $k_\textrm{e}$ and $k_\textrm{o}$ are independent of spin. This reveals a clear distinction for distinguishable collisions, as each collision gives rise to a loss of a single molecule from each respective spin state. Summing the terms in Eq.\ (\ref{eq:distingusihable1}) while assuming equal population for each spin state, $n_1 =  n_2 = n_3= \frac{n}{3}$, results in a two-body loss rate
 \begin{equation}
    \dot{n} = - \left( \frac{4}{3} k_\textrm{e}  + \frac{2}{3} k_\textrm{o} \right)n^2 .
        \label{eq:distingusihablefinal}
\end{equation}
Following the example above, the two-body loss rate for collisions between $q$ equally populated spin states scales according to the simple analytical form,
 \begin{equation}
    \dot{n} = - \left( \frac{q + 1}{q} k_\textrm{e}  + \frac{q-1}{q} k_\textrm{o} \right)  n^2 .
        \label{eq:distingusihableanalytical}
\end{equation}

As a consequence of Eq.\ (\ref{eq:distingusihableanalytical}), the rate coefficient among increasingly large distributions of spin states will simply reduce to the sum of $k_\textrm{e}$ and $k_\textrm{o}$. Therefore, we can use experiments with different numbers of populated states and temperatures to disentangle the contributions to the total loss rate. In particular, we experimentally access five different conditions: unpolarized $N=1$ (12 states) at three temperatures; unpolarized $N=0$, $G=1$ (3 states); and polarized $N=0$ (1 state).

Several physical mechanisms can cause loss. YO is strongly bound, by 7.3 eV \cite{ackermann1964, ackermann1974},
so the bond-exchange reaction, $\mathrm{YO} + \mathrm{YO} \rightarrow \mathrm{Y_{2}} + \mathrm{O_{2}}$, is energetically forbidden \cite{brix1954, fang2000}. It is not known whether the trimer-formation channels $\mathrm{Y_2O} + \mathrm{O}$ and $\mathrm{YO_2} + \mathrm{Y}$ are energetically accessible, but loss could also be caused by photo-excitation of the collision complex or three-body collisions. Any of these processes is likely to be enhanced by long-lived ``sticky'' collisions \cite{mayle2012, mayle2013, Christianen:density:2019, Christianen:laser:2019}.
Molecules prepared in excited states can also be lost by rotational and/or hyperfine relaxation.

To capture the multiple potential causes of loss, we use a QDT approach. The loss is modeled by a short-range absorbing boundary condition, regardless of the physical mechanism. This is controlled by the loss parameter $y$, ranging from no loss at $y=0$ to complete ``universal'' short-range loss at $y=1$, and a short-range phase shift $\delta^\textrm{s}$ \cite{Idziaszek:PRL:2010, Frye:2015, Gao:2008}.
This approach has been successful in modeling loss in similar systems \cite{Ospelkaus:react:2010, Gregory_2019}.
We calculate cross sections and thermally averaged two-body loss rate coefficients \cite{Frye:2015} using analytic QDT functions \cite{Gao:C6:1998, Gao:AQDTroutines}, including all partial waves needed for convergence.

The total density loss rate $\dot{n}_{\mathrm{mol}}$ is defined as
\begin{equation}
    \dot{n}_{\mathrm{mol}} = -\alpha n_{\mathrm{mol}} - (2k_2) n_{\mathrm{mol}}^2 ,
        \label{eq:loss_rate}
\end{equation}
where $\alpha$  and $k_2$ are the one-body and two-body loss rate coefficients. The density is determined using the molecule number 
 and $V_{\mathrm{eff}}$ which is the effective volume defined as $V_{\mathrm{eff}}$ = $8\sqrt{\pi^3} \sigma_x \sigma_y \sigma_z$ where $\sigma_i$ is the rms width of the thermal distribution along each direction in the XODT with $i$ = $\{x,y,z\}$. The rms widths are determined by the equipartition theorem $m \omega^{2}_i\sigma^{2}_i$  = $k_\textrm{B}T_{i}$, where $k_\textrm{B}$ is Boltzmann's constant, $T_i$ is the temperature, and $m$ is the mass of YO. The temperature of the molecules is measured by time-of-flight expansion in the radial $\{x,y\}$ direction and the axial $\{z\}$ direction. The trapping frequencies $\omega_i$ are determined by parametrically heating the molecules out of the XODT, which is formed by two independent 1064-nm laser beams intersecting at 45$^\circ$. 

\begin{figure}[t!]
\includegraphics[width= 1 \linewidth]{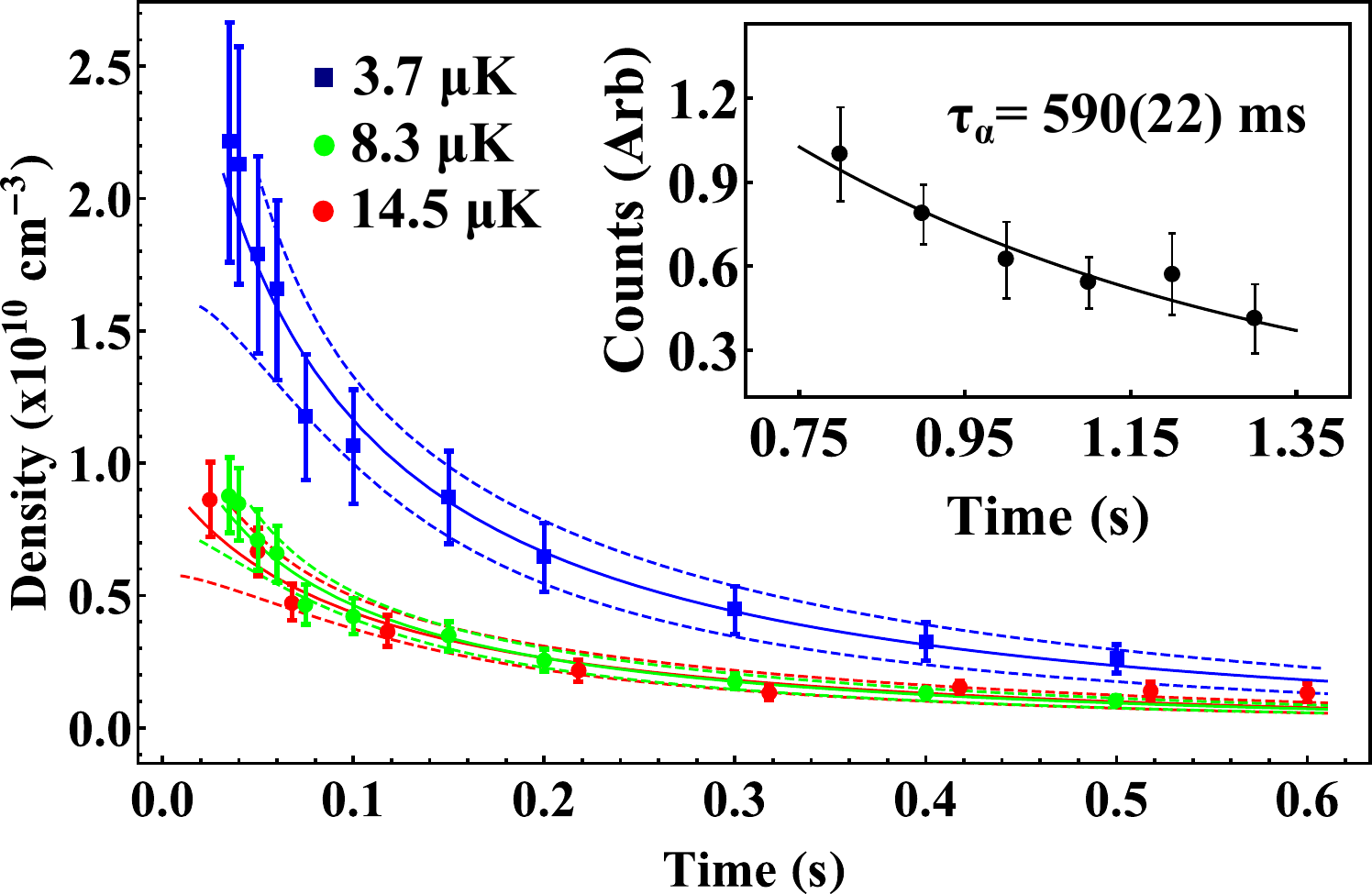}
\caption{
\label{fig:N = 1} Collisional loss in an incoherent spin mixture of YO molecules in the $N = 1$ rotational manifold. The collisional loss rate is measured at three different average temperatures of 3.7(5) $\mu$K (solid blue squares) and, 8.3(1.1), 14.5(1.7) $\mu$K (solid green and solid red circles) which are below and above the p-wave centrifugal barrier height, respectively. The solid lines represent the fit of the data points to the density loss rate according to Eq.\ (\ref{eq:loss_rate}), and the dashed lines represent the 95\% confidence interval bands from the fits.  By tuning the temperature below and above the p-wave barrier height we determine two-body loss rates of $k_2 (3.7\, \mu$K$) = 2.2(0.6) \times 10^{-10}$ cm$^3$\,s$^{-1}$, $k_2 (8.3\, \mu$K$) = 5.6(1.8) \times 10^{-10}$ cm$^3$\,s$^{-1}$ and $k_2 (14.5\, \mu$K$) = 4.9(1.4) \times 10^{-10}$ cm$^3$\,s$^{-1}$. To quantify the fitting results we find a reduced chi-squared of 0.2, 0.3 and 1.4 for the respective fits. The inset shows the one-body lifetime $\tau_\alpha$ = 590(22) ms.
}
\end{figure}

To load our XODT optimally, we apply 40 ms of GMC while ramping on the 1064-nm beams.  Collisional studies of an incoherent mixture of molecules in different internal states of the $N = 1$ rotational manifold follow naturally after GMC, as the molecules are spread over 12 spin states during the overlap with the XODT beams. The number of molecules left in the XODT as a function of hold time is then measured by applying a 1.5 ms readout pulse. The total loss rate for three different temperatures is shown in Fig.\ \ref{fig:N = 1}. The first trace (solid blue squares) and the second trace (solid green circles) are taken at a trap depth of 85 $\mu$K. The average temperatures of the molecules, defined as $T_{\mathrm{avg}} = T^{2/3}_{\mathrm{radial}} \times T^{1/3}_{\mathrm{axial}}$, are 3.7(5) $\mu$K and 8.3(1.1) $\mu$K respectively, which are below and above the p-wave barrier height of 4.2 $\mu$K. The third trace (solid red circles) has a higher trap depth of 150 $\mu$K. The average temperature of the molecules is 14.5(1.7) $\mu$K, which is well above the p-wave but below the d-wave barrier height of 22 $\mu$K. The number of trapped molecules is $\sim$5000 in all cases with a 40\% capture efficiency from the blue-detuned MOT. The one-body lifetime of 590(22) ms is measured at long hold times (Fig.\ \ref{fig:N = 1} inset). 

Using Eq.\ (\ref{eq:loss_rate}), a weighted fit for the curves in Fig.\ \ref{fig:N = 1} is performed while fixing $\alpha$. A change in temperature with time is not measured within experimental error, so we fix the trapping volume $V_\textrm{eff}$. The experimental results are $k_2 (3.7\, \mu$K$) = 2.2(0.6) \times 10^{-10}$ cm$^3$\,s$^{-1}$, $k_2 (8.3\, \mu$K$) = 5.6(1.8) \times 10^{-10}$ cm$^3$\,s$^{-1}$ and $k_2 (14.5\, \mu$K$) = 4.9(1.4) \times 10^{-10}$ cm$^3$\,s$^{-1}$. The results of our universal-loss QDT calculations are $k_2 (3.7\, \mu \textrm{K})=1.3\times 10^{-10}$ cm$^3$\,s$^{-1}$ and $k_2 (14.5\, \mu \textrm{K})=1.5\times 10^{-10}$ cm$^3$\,s$^{-1}$. At the lower temperature, there is a fair agreement between theory and experiment. Both show an increased loss at the higher temperature, although it is more pronounced in the experiment. Such a feature could arise if there is a resonant enhancement for higher partial waves.

\begin{figure*}[t!]
\centering
\includegraphics[width= 1 \textwidth]{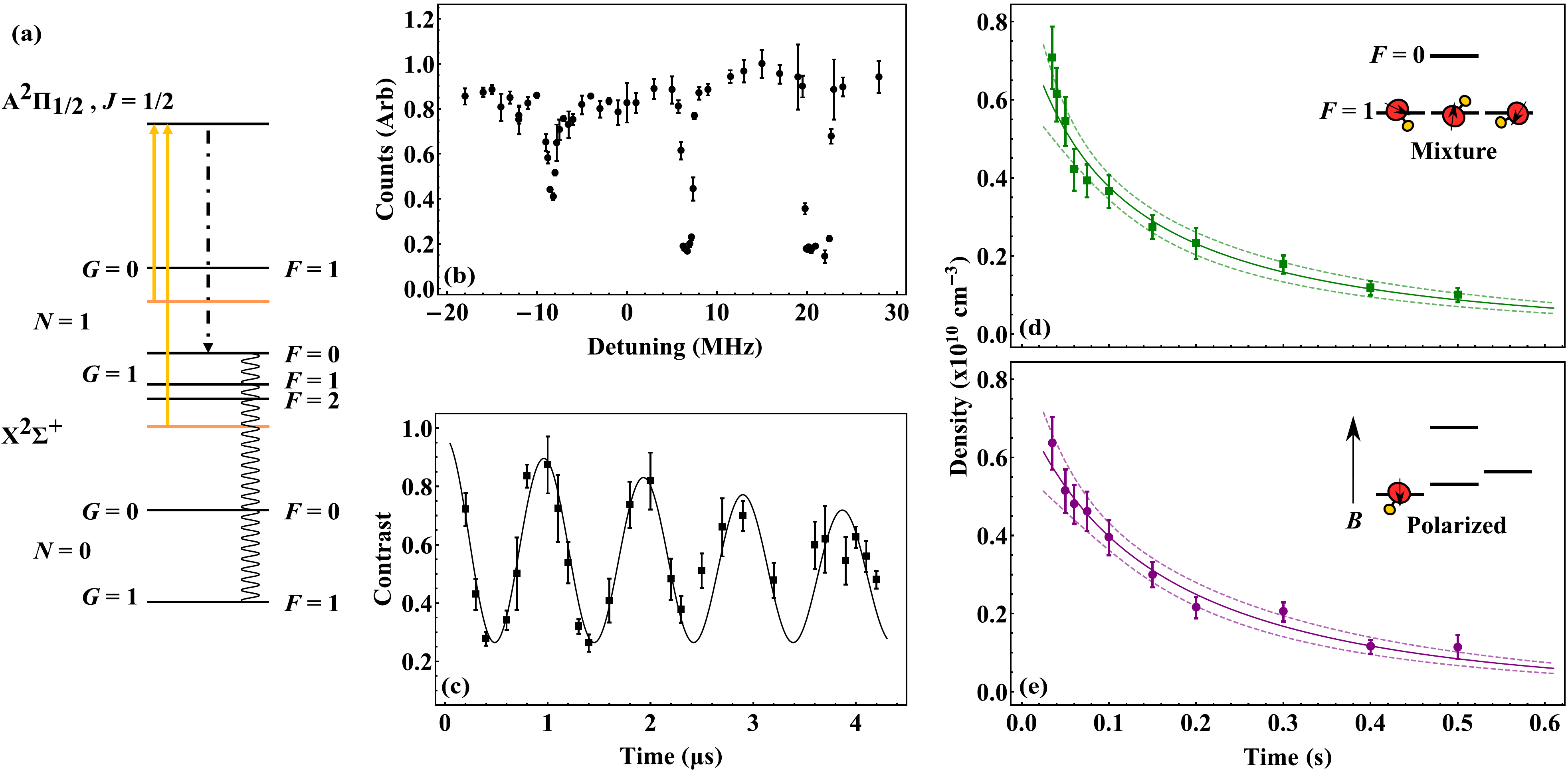}
\caption{
\label{fig:optical Pumping} (a) The optical pumping sequence to transfer molecules in $N = 1$ into a single hyperfine level $G = 1, F = 0$. This is achieved by applying two beams blue-detuned from $N = 1, G = 0, F = 1$  and $N = 1, G = 1, F = 1,2$. Microwave fields are then swept across the transition $N = 1, G = 1, F = 0$ $\rightarrow$ $N = 0, G = 1, F = 1$ to transfer the molecules into the absolute ground state. (b)  Microwave spectroscopy of the transition with an applied magnetic field, showing splitting of the hyperfine sublevels in $N = 0, G = 1, F = 1$. (c) Rabi oscillations between $N$ = 1 and $N$ = 0. (d) Collisional loss of YO molecules prepared in an equal mixture of spin states in $N = 0, G = 1, F = 1$ (solid green squares). (e) Collisional loss in a single spin-polarized state $N = 0, G = 1, F = 1, m_{F}= -1$ (solid purple circles). The solid lines represent the fitting of the data points to the rate equation given in Eq.\ \ref{eq:loss_rate}, and the dashed lines represent the 95\% confidence interval bands from the fits. To quantify the fitting results we find a reduced chi-squared of 0.8 and 0.7 for the respective fits. We determine rate coefficients of $k_\textrm{mix}= 5.4(1.5) \times 10^{-10}$ cm$^3$\,s$^{-1}$ and $k_\textrm{pol}= 3.2(1.0) \times 10^{-10}$ cm$^3$\,s$^{-1}$ at 5.8 $\mu$K in the $N$ = 0 ground state.
} 
\end{figure*}\textbf{}
To study collisions in a spin-polarized case, the molecules are next prepared in the absolute rovibrational ground state, the $N = 0, G = 1, F = 1$ manifold,  both in an incoherent mixture of the $m_F$ sublevels and spin-polarized in $m_F = -1$. The transfer of the molecules into the absolute ground state begins by optically pumping into the single quantum state $N = 1, G = 1, F = 0$ by applying two laser tones blue detuned from $N = 1, G = 0, F = 1$ and $N = 1, G = 1, F = 2$. This process is shown diagrammatically in Fig.\ \ref{fig:optical Pumping}(a). A magnetic field of  $\sim$16 G is applied along $z$ to split the degeneracy of the $N = 0, G = 1, F = 1$ hyperfine sublevels; applying a microwave sweep transfers the molecules into a particular sublevel of the absolute ground state. To produce an incoherent mixture, no magnetic field is applied during the sweep. The resonant frequency  that addresses the transition $N = 1, G = 1, F = 0$ $\rightarrow$ $N = 0, G = 1, F = 1$ transition is 23.282405 GHz~\cite{microwave}. 

Microwave spectroscopy is shown in Fig.\ \ref{fig:optical Pumping}(b) with an applied magnetic field. A clear separation of the three sublevels is observed. As the microwave frequency is red-detuned at zero field, no mixing of the population between $N = 1, G = 1, F = 1, 2$ and $N = 0, G = 1, F = 1$ is observed, suggesting that the optical pumping to $N = 1, G = 1, F = 0$ has high fidelity ($>$95\%).  Figure \ref{fig:optical Pumping}(c) shows Rabi oscillations between the $N$ = 1 and $N$ = 0 rotational levels in free space. Rabi rates of   $\sim$1 MHz are achieved. Fast dephasing of the oscillations is measured with a coherence time of 8.8 $\mu$s. We attribute this to an inhomogeneous microwave field inside our metal vacuum chamber. This limits our coherent-transfer fidelity to the desired sublevel(s) in $N$ = 0 to 70\%. After removing all unwanted population,  the transferred population in $N = 0$ is measured as a function of time, recovering the associated collision loss rate. 

The collisional loss rates for both an incoherent mixture (solid green squares) and spin-polarized YO molecules (solid purple circles) are shown in Figs.\ \ref{fig:optical Pumping}(d) and \ref{fig:optical Pumping}(e), all under an XODT trap depth of 85 $\mu$K. Due to optical pumping, the molecules are heated to an average temperature of 5.8(5) $\mu$K, just above the p-wave barrier height. Fitting to Eq.\ (\ref{eq:loss_rate}) for the total loss rates, we determine $k_\textrm{mix}= 5.4(1.5) \times 10^{-10}$ cm$^3$\,s$^{-1}$ and $k_\textrm{pol}= 3.2(1.0) \times 10^{-10}$ cm$^3$\,s$^{-1}$. The corresponding calculated values for universal loss are $k_\textrm{mix} = 1.3 \times 10^{-10}$ cm$^3$\,s$^{-1}$ and $k_\textrm{pol} = 1.3 \times 10^{-10}$ cm$^3$\,s$^{-1}$. The experimental $k_\textrm{mix}$ is larger than the theory; this again could be a sign of resonant enhancement, most likely in p-wave scattering. This discrepancy may also arise from the density determination, which is derived from the molecule number, temperature, and trap frequencies. 

\begin{figure}[t!]
\centering
\includegraphics[width= 1 \linewidth]{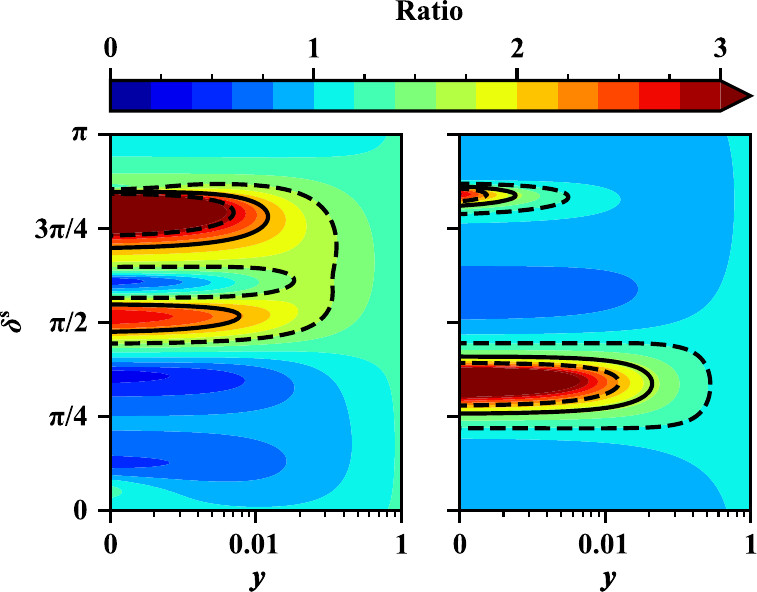}
\caption{
\label{fig:theory_contour} Contour plots of the calculated ratios $k_2(14.5\, \mu$K$)/k_2(3.7\, \mu$K$)$ for $N=1$ (left), and $k_\textrm{mix}/k_\textrm{pol}$ for $N=0$ (right). Experimentally measured ratios are shown as solid black lines, with uncertainties indicated by dashed lines. The vertical and horizontal axes represent the short-range phase shift $\delta^s$ and the loss parameter $y$ respectively.
}
\end{figure}

Our results for both rotational states suggest the presence of resonances, particularly in N = 1.
We therefore focus on the ratios of rates, between high and low temperature for $N=1$ and between spin mixtures and polarized samples for $N=0$. Resonances cannot exist for universal loss ($y=1$) since there is no reflection to produce interference. However, if $y<1$, resonances occur when the short-range phase shift $\delta^\textrm{s}$ produces constructive interference.
Our QDT model represents only shape resonances explicitly, but broad Feshbach resonances will give the same observable effects ~\cite{Frye:2015, Frye:triatomic-complexes:2021}.
Figure \ref{fig:theory_contour} shows the ratios from the QDT model and compares them with experiment.
For $N=1$, the experimental ratio can occur in two regions with $y$ below 0.25 and $\delta^\textrm{s}$ just below $5\pi/8$ and $7\pi/8$; these correspond to d- and f-wave shape resonances, respectively \cite{Gao:2000, Frye:2015}. Notably, a p-wave shape resonance is incapable of reproducing the experimental ratio. For $N=0$ the region of agreement is again at low $y$, but now near $\delta^\textrm{s}=3\pi/8$, corresponding to a p-wave shape resonance. This provides valuable insight into the dynamics of this new ultracold system.

In conclusion, we report a direct measurement of loss rates in the lowest partial waves for laser-cooled molecules. Our results suggest that there is incomplete loss at short range, and that resonances in higher partial waves play an important role. With an understanding of the collisional loss rates and with the large permanent dipole moment in YO molecules, collisional shielding can be achieved by the application of either dc or microwave fields to significantly enhance the ratio of elastic to inelastic collision rates. Evaporative cooling will then be applied with the aim of achieving a quantum degenerate gas of laser-cooled molecules. 

\begin{acknowledgments}
 We thank S. Scheidegger and J. Higgins for careful reading and useful feedback on the manuscript, and M. Tomza for helpful discussions. Funding support for this work is provided by ARO MURI, AFOSR MURI, NIST, and NSF PHY-2317149. The theoretical work was supported by the U.K. Engineering and Physical Sciences Research Council (EPSRC) Grants No.\ EP/W00299X/1, EP/P01058X/1, and EP/V011677/1.
\end{acknowledgments}
\nocite{*}

\bibliography{main.bib}% Produces the bibliography via BibTeX.

\begin{thebibliography}{61}%
\makeatletter
\providecommand \@ifxundefined [1]{%
 \@ifx{#1\undefined}
}%
\providecommand \@ifnum [1]{%
 \ifnum #1\expandafter \@firstoftwo
 \else \expandafter \@secondoftwo
 \fi
}%
\providecommand \@ifx [1]{%
 \ifx #1\expandafter \@firstoftwo
 \else \expandafter \@secondoftwo
 \fi
}%
\providecommand \natexlab [1]{#1}%
\providecommand \enquote  [1]{``#1''}%
\providecommand \bibnamefont  [1]{#1}%
\providecommand \bibfnamefont [1]{#1}%
\providecommand \citenamefont [1]{#1}%
\providecommand \href@noop [0]{\@secondoftwo}%
\providecommand \href [0]{\begingroup \@sanitize@url \@href}%
\providecommand \@href[1]{\@@startlink{#1}\@@href}%
\providecommand \@@href[1]{\endgroup#1\@@endlink}%
\providecommand \@sanitize@url [0]{\catcode `\\12\catcode `\$12\catcode `\&12\catcode `\#12\catcode `\^12\catcode `\_12\catcode `\%12\relax}%
\providecommand \@@startlink[1]{}%
\providecommand \@@endlink[0]{}%
\providecommand \url  [0]{\begingroup\@sanitize@url \@url }%
\providecommand \@url [1]{\endgroup\@href {#1}{\urlprefix }}%
\providecommand \urlprefix  [0]{URL }%
\providecommand \Eprint [0]{\href }%
\providecommand \doibase [0]{https://doi.org/}%
\providecommand \selectlanguage [0]{\@gobble}%
\providecommand \bibinfo  [0]{\@secondoftwo}%
\providecommand \bibfield  [0]{\@secondoftwo}%
\providecommand \translation [1]{[#1]}%
\providecommand \BibitemOpen [0]{}%
\providecommand \bibitemStop [0]{}%
\providecommand \bibitemNoStop [0]{.\EOS\space}%
\providecommand \EOS [0]{\spacefactor3000\relax}%
\providecommand \BibitemShut  [1]{\csname bibitem#1\endcsname}%
\let\auto@bib@innerbib\@empty
%</preamble>
\bibitem [{\citenamefont {De~Marco}\ \emph {et~al.}(2019)\citenamefont {De~Marco}, \citenamefont {Valtolina}, \citenamefont {Matsuda}, \citenamefont {Tobias}, \citenamefont {Covey},\ and\ \citenamefont {Ye}}]{Marco2019}%
  \BibitemOpen
  \bibfield  {author} {\bibinfo {author} {\bibfnamefont {L.}~\bibnamefont {De~Marco}}, \bibinfo {author} {\bibfnamefont {G.}~\bibnamefont {Valtolina}}, \bibinfo {author} {\bibfnamefont {K.}~\bibnamefont {Matsuda}}, \bibinfo {author} {\bibfnamefont {W.~G.}\ \bibnamefont {Tobias}}, \bibinfo {author} {\bibfnamefont {J.~P.}\ \bibnamefont {Covey}},\ and\ \bibinfo {author} {\bibfnamefont {J.}~\bibnamefont {Ye}},\ }\bibfield  {title} {\bibinfo {title} {{A Degenerate Fermi Gas of Polar Molecules}},\ }\href {https://doi.org/10.1126/science.aau7230} {\bibfield  {journal} {\bibinfo  {journal} {Science}\ }\textbf {\bibinfo {volume} {363}},\ \bibinfo {pages} {853} (\bibinfo {year} {2019})}\BibitemShut {NoStop}%
\bibitem [{\citenamefont {Valtolina}\ \emph {et~al.}(2020)\citenamefont {Valtolina}, \citenamefont {Matsuda}, \citenamefont {Tobias}, \citenamefont {Li}, \citenamefont {De~Marco},\ and\ \citenamefont {Ye}}]{Valtolina2020}%
  \BibitemOpen
  \bibfield  {author} {\bibinfo {author} {\bibfnamefont {G.}~\bibnamefont {Valtolina}}, \bibinfo {author} {\bibfnamefont {K.}~\bibnamefont {Matsuda}}, \bibinfo {author} {\bibfnamefont {W.~G.}\ \bibnamefont {Tobias}}, \bibinfo {author} {\bibfnamefont {J.-R.}\ \bibnamefont {Li}}, \bibinfo {author} {\bibfnamefont {L.}~\bibnamefont {De~Marco}},\ and\ \bibinfo {author} {\bibfnamefont {J.}~\bibnamefont {Ye}},\ }\bibfield  {title} {\bibinfo {title} {{Dipolar Evaporation of Reactive Molecules to Below the Fermi Temperature}},\ }\href@noop {} {\bibfield  {journal} {\bibinfo  {journal} {Nature}\ }\textbf {\bibinfo {volume} {588}},\ \bibinfo {pages} {239 } (\bibinfo {year} {2020})}\BibitemShut {NoStop}%
\bibitem [{\citenamefont {Schindewolf}\ \emph {et~al.}(2022)\citenamefont {Schindewolf}, \citenamefont {Bause}, \citenamefont {Chen}, \citenamefont {Duda}, \citenamefont {Bloch},\ and\ \citenamefont {Luo}}]{Schindewolf2022}%
  \BibitemOpen
  \bibfield  {author} {\bibinfo {author} {\bibfnamefont {A.}~\bibnamefont {Schindewolf}}, \bibinfo {author} {\bibfnamefont {R.}~\bibnamefont {Bause}}, \bibinfo {author} {\bibfnamefont {X.-Y.}\ \bibnamefont {Chen}}, \bibinfo {author} {\bibfnamefont {T.}~\bibnamefont {Duda}, \bibfnamefont {M.~Karman}}, \bibinfo {author} {\bibfnamefont {I.}~\bibnamefont {Bloch}},\ and\ \bibinfo {author} {\bibfnamefont {X.-Y.}\ \bibnamefont {Luo}},\ }\bibfield  {title} {\bibinfo {title} {{Evaporation of Microwave-shielded Polar Molecules to Quantum Degeneracy}},\ }\href {https://doi.org/10.1038/s41586-022-04900-0} {\bibfield  {journal} {\bibinfo  {journal} {Nature}\ }\textbf {\bibinfo {volume} {607}},\ \bibinfo {pages} {677} (\bibinfo {year} {2022})}\BibitemShut {NoStop}%
\bibitem [{\citenamefont {Cao}\ \emph {et~al.}(2023)\citenamefont {Cao}, \citenamefont {Yang}, \citenamefont {Su}, \citenamefont {Wang}, \citenamefont {Rui}, \citenamefont {Zhao},\ and\ \citenamefont {Pan}}]{Pan2022}%
  \BibitemOpen
  \bibfield  {author} {\bibinfo {author} {\bibfnamefont {J.}~\bibnamefont {Cao}}, \bibinfo {author} {\bibfnamefont {H.}~\bibnamefont {Yang}}, \bibinfo {author} {\bibfnamefont {Z.}~\bibnamefont {Su}}, \bibinfo {author} {\bibfnamefont {X.-Y.}\ \bibnamefont {Wang}}, \bibinfo {author} {\bibfnamefont {J.}~\bibnamefont {Rui}}, \bibinfo {author} {\bibfnamefont {B.}~\bibnamefont {Zhao}},\ and\ \bibinfo {author} {\bibfnamefont {J.-W.}\ \bibnamefont {Pan}},\ }\bibfield  {title} {\bibinfo {title} {{Preparation of a Quantum Degenerate Mixture of $^{23}\mathrm{Na}^{40}\mathrm{K}$ Molecules and $^{40}\mathrm{K}$ Atoms}},\ }\href {https://doi.org/10.1103/PhysRevA.107.013307} {\bibfield  {journal} {\bibinfo  {journal} {Phys. Rev. A}\ }\textbf {\bibinfo {volume} {107}},\ \bibinfo {pages} {013307} (\bibinfo {year} {2023})}\BibitemShut {NoStop}%
\bibitem [{\citenamefont {Bigagli}\ \emph {et~al.}(2024)\citenamefont {Bigagli}, \citenamefont {Yuan}, \citenamefont {Zhang}, \citenamefont {Bulatovic}, \citenamefont {Karman}, \citenamefont {Stevenson},\ and\ \citenamefont {Will}}]{bigagli2023observation}%
  \BibitemOpen
  \bibfield  {author} {\bibinfo {author} {\bibfnamefont {N.}~\bibnamefont {Bigagli}}, \bibinfo {author} {\bibfnamefont {W.}~\bibnamefont {Yuan}}, \bibinfo {author} {\bibfnamefont {S.}~\bibnamefont {Zhang}}, \bibinfo {author} {\bibfnamefont {B.}~\bibnamefont {Bulatovic}}, \bibinfo {author} {\bibfnamefont {T.}~\bibnamefont {Karman}}, \bibinfo {author} {\bibfnamefont {I.}~\bibnamefont {Stevenson}},\ and\ \bibinfo {author} {\bibfnamefont {S.}~\bibnamefont {Will}},\ }\bibfield  {title} {\bibinfo {title} {{Observation of {Bose--Einstein} Condensation of Dipolar Molecules}},\ }\href@noop {} {\bibfield  {journal} {\bibinfo  {journal} {Nature}\ }\textbf {\bibinfo {volume} {631}},\ \bibinfo {pages} {289} (\bibinfo {year} {2024})}\BibitemShut {NoStop}%
\bibitem [{\citenamefont {Li}\ \emph {et~al.}(2023{\natexlab{a}})\citenamefont {Li}, \citenamefont {Matsuda}, \citenamefont {Miller}, \citenamefont {Carroll}, \citenamefont {Tobias}, \citenamefont {Higgins},\ and\ \citenamefont {Ye}}]{Li2023}%
  \BibitemOpen
  \bibfield  {author} {\bibinfo {author} {\bibfnamefont {J.-R.}\ \bibnamefont {Li}}, \bibinfo {author} {\bibfnamefont {K.}~\bibnamefont {Matsuda}}, \bibinfo {author} {\bibfnamefont {C.}~\bibnamefont {Miller}}, \bibinfo {author} {\bibfnamefont {A.~N.}\ \bibnamefont {Carroll}}, \bibinfo {author} {\bibfnamefont {W.~G.}\ \bibnamefont {Tobias}}, \bibinfo {author} {\bibfnamefont {J.~S.}\ \bibnamefont {Higgins}},\ and\ \bibinfo {author} {\bibfnamefont {J.}~\bibnamefont {Ye}},\ }\bibfield  {title} {\bibinfo {title} {{Tunable Itinerant Spin Dynamics with Polar Molecules}},\ }\href@noop {} {\bibfield  {journal} {\bibinfo  {journal} {Nature}\ }\textbf {\bibinfo {volume} {614}},\ \bibinfo {pages} {70} (\bibinfo {year} {2023}{\natexlab{a}})}\BibitemShut {NoStop}%
\bibitem [{\citenamefont {B\"uchler}\ \emph {et~al.}(2007)\citenamefont {B\"uchler}, \citenamefont {Demler}, \citenamefont {Lukin}, \citenamefont {Micheli}, \citenamefont {Prokof'ev}, \citenamefont {Pupillo},\ and\ \citenamefont {Zoller}}]{Correlated}%
  \BibitemOpen
  \bibfield  {author} {\bibinfo {author} {\bibfnamefont {H.~P.}\ \bibnamefont {B\"uchler}}, \bibinfo {author} {\bibfnamefont {E.}~\bibnamefont {Demler}}, \bibinfo {author} {\bibfnamefont {M.}~\bibnamefont {Lukin}}, \bibinfo {author} {\bibfnamefont {A.}~\bibnamefont {Micheli}}, \bibinfo {author} {\bibfnamefont {N.}~\bibnamefont {Prokof'ev}}, \bibinfo {author} {\bibfnamefont {G.}~\bibnamefont {Pupillo}},\ and\ \bibinfo {author} {\bibfnamefont {P.}~\bibnamefont {Zoller}},\ }\bibfield  {title} {\bibinfo {title} {{Strongly Correlated 2D Quantum Phases with Cold Polar Molecules: Controlling the Shape of the Interaction Potential}},\ }\href {https://doi.org/10.1103/PhysRevLett.98.060404} {\bibfield  {journal} {\bibinfo  {journal} {Phys. Rev. Lett.}\ }\textbf {\bibinfo {volume} {98}},\ \bibinfo {pages} {060404} (\bibinfo {year} {2007})}\BibitemShut {NoStop}%
\bibitem [{\citenamefont {Defenu}\ \emph {et~al.}(2023)\citenamefont {Defenu}, \citenamefont {Donner}, \citenamefont {Macr\`{\i}}, \citenamefont {Pagano}, \citenamefont {Ruffo},\ and\ \citenamefont {Trombettoni}}]{longrange}%
  \BibitemOpen
  \bibfield  {author} {\bibinfo {author} {\bibfnamefont {N.}~\bibnamefont {Defenu}}, \bibinfo {author} {\bibfnamefont {T.}~\bibnamefont {Donner}}, \bibinfo {author} {\bibfnamefont {T.}~\bibnamefont {Macr\`{\i}}}, \bibinfo {author} {\bibfnamefont {G.}~\bibnamefont {Pagano}}, \bibinfo {author} {\bibfnamefont {S.}~\bibnamefont {Ruffo}},\ and\ \bibinfo {author} {\bibfnamefont {A.}~\bibnamefont {Trombettoni}},\ }\bibfield  {title} {\bibinfo {title} {{Long-range Interacting Quantum Systems}},\ }\href {https://doi.org/10.1103/RevModPhys.95.035002} {\bibfield  {journal} {\bibinfo  {journal} {Rev. Mod. Phys.}\ }\textbf {\bibinfo {volume} {95}},\ \bibinfo {pages} {035002} (\bibinfo {year} {2023})}\BibitemShut {NoStop}%
\bibitem [{\citenamefont {Pollet}\ \emph {et~al.}(2010)\citenamefont {Pollet}, \citenamefont {Picon}, \citenamefont {B\"uchler},\ and\ \citenamefont {Troyer}}]{Triangle}%
  \BibitemOpen
  \bibfield  {author} {\bibinfo {author} {\bibfnamefont {L.}~\bibnamefont {Pollet}}, \bibinfo {author} {\bibfnamefont {J.~D.}\ \bibnamefont {Picon}}, \bibinfo {author} {\bibfnamefont {H.~P.}\ \bibnamefont {B\"uchler}},\ and\ \bibinfo {author} {\bibfnamefont {M.}~\bibnamefont {Troyer}},\ }\bibfield  {title} {\bibinfo {title} {{Supersolid Phase with Cold Polar Molecules on a Triangular Lattice}},\ }\href {https://doi.org/10.1103/PhysRevLett.104.125302} {\bibfield  {journal} {\bibinfo  {journal} {Phys. Rev. Lett.}\ }\textbf {\bibinfo {volume} {104}},\ \bibinfo {pages} {125302} (\bibinfo {year} {2010})}\BibitemShut {NoStop}%
\bibitem [{\citenamefont {G\'oral}\ \emph {et~al.}(2002)\citenamefont {G\'oral}, \citenamefont {Santos},\ and\ \citenamefont {Lewenstein}}]{dipolarlattice}%
  \BibitemOpen
  \bibfield  {author} {\bibinfo {author} {\bibfnamefont {K.}~\bibnamefont {G\'oral}}, \bibinfo {author} {\bibfnamefont {L.}~\bibnamefont {Santos}},\ and\ \bibinfo {author} {\bibfnamefont {M.}~\bibnamefont {Lewenstein}},\ }\bibfield  {title} {\bibinfo {title} {{Quantum Phases of Dipolar Bosons in Optical Lattices}},\ }\href {https://doi.org/10.1103/PhysRevLett.88.170406} {\bibfield  {journal} {\bibinfo  {journal} {Phys. Rev. Lett.}\ }\textbf {\bibinfo {volume} {88}},\ \bibinfo {pages} {170406} (\bibinfo {year} {2002})}\BibitemShut {NoStop}%
\bibitem [{\citenamefont {Schmidt}\ \emph {et~al.}(2022)\citenamefont {Schmidt}, \citenamefont {Lassabli\`ere}, \citenamefont {Qu\'em\'ener},\ and\ \citenamefont {Langen}}]{bound}%
  \BibitemOpen
  \bibfield  {author} {\bibinfo {author} {\bibfnamefont {M.}~\bibnamefont {Schmidt}}, \bibinfo {author} {\bibfnamefont {L.}~\bibnamefont {Lassabli\`ere}}, \bibinfo {author} {\bibfnamefont {G.}~\bibnamefont {Qu\'em\'ener}},\ and\ \bibinfo {author} {\bibfnamefont {T.}~\bibnamefont {Langen}},\ }\bibfield  {title} {\bibinfo {title} {{Self-bound Dipolar Droplets and Supersolids in Molecular Bose-Einstein Condensates}},\ }\href {https://doi.org/10.1103/PhysRevResearch.4.013235} {\bibfield  {journal} {\bibinfo  {journal} {Phys. Rev. Res.}\ }\textbf {\bibinfo {volume} {4}},\ \bibinfo {pages} {013235} (\bibinfo {year} {2022})}\BibitemShut {NoStop}%
\bibitem [{\citenamefont {Norrgard}\ \emph {et~al.}(2016)\citenamefont {Norrgard}, \citenamefont {McCarron}, \citenamefont {Steinecker}, \citenamefont {Tarbutt},\ and\ \citenamefont {DeMille}}]{eric}%
  \BibitemOpen
  \bibfield  {author} {\bibinfo {author} {\bibfnamefont {E.~B.}\ \bibnamefont {Norrgard}}, \bibinfo {author} {\bibfnamefont {D.~J.}\ \bibnamefont {McCarron}}, \bibinfo {author} {\bibfnamefont {M.~H.}\ \bibnamefont {Steinecker}}, \bibinfo {author} {\bibfnamefont {M.~R.}\ \bibnamefont {Tarbutt}},\ and\ \bibinfo {author} {\bibfnamefont {D.}~\bibnamefont {DeMille}},\ }\bibfield  {title} {\bibinfo {title} {{Submillikelvin Dipolar Molecules in a Radio-Frequency Magneto-Optical Trap}},\ }\href {https://doi.org/10.1103/PhysRevLett.116.063004} {\bibfield  {journal} {\bibinfo  {journal} {Phys. Rev. Lett.}\ }\textbf {\bibinfo {volume} {116}},\ \bibinfo {pages} {063004} (\bibinfo {year} {2016})}\BibitemShut {NoStop}%
\bibitem [{\citenamefont {Truppe}\ \emph {et~al.}(2017)\citenamefont {Truppe}, \citenamefont {Williams}, \citenamefont {Hambach}, \citenamefont {Caldwell}, \citenamefont {Fitch}, \citenamefont {Hinds}, \citenamefont {Sauer},\ and\ \citenamefont {Tarbutt}}]{Truppe_2017}%
  \BibitemOpen
  \bibfield  {author} {\bibinfo {author} {\bibfnamefont {S.}~\bibnamefont {Truppe}}, \bibinfo {author} {\bibfnamefont {H.~J.}\ \bibnamefont {Williams}}, \bibinfo {author} {\bibfnamefont {M.}~\bibnamefont {Hambach}}, \bibinfo {author} {\bibfnamefont {L.}~\bibnamefont {Caldwell}}, \bibinfo {author} {\bibfnamefont {N.~J.}\ \bibnamefont {Fitch}}, \bibinfo {author} {\bibfnamefont {E.~A.}\ \bibnamefont {Hinds}}, \bibinfo {author} {\bibfnamefont {B.~E.}\ \bibnamefont {Sauer}},\ and\ \bibinfo {author} {\bibfnamefont {M.~R.}\ \bibnamefont {Tarbutt}},\ }\bibfield  {title} {\bibinfo {title} {{Molecules Cooled Below the Doppler Limit}},\ }\href {https://doi.org/10.1038/nphys4241} {\bibfield  {journal} {\bibinfo  {journal} {Nature Physics}\ }\textbf {\bibinfo {volume} {13}},\ \bibinfo {pages} {1173–1176} (\bibinfo {year} {2017})}\BibitemShut {NoStop}%
\bibitem [{\citenamefont {Collopy}\ \emph {et~al.}(2018)\citenamefont {Collopy}, \citenamefont {Ding}, \citenamefont {Wu}, \citenamefont {Finneran}, \citenamefont {Anderegg}, \citenamefont {Augenbraun}, \citenamefont {Doyle},\ and\ \citenamefont {Ye}}]{3dYO}%
  \BibitemOpen
  \bibfield  {author} {\bibinfo {author} {\bibfnamefont {A.~L.}\ \bibnamefont {Collopy}}, \bibinfo {author} {\bibfnamefont {S.}~\bibnamefont {Ding}}, \bibinfo {author} {\bibfnamefont {Y.}~\bibnamefont {Wu}}, \bibinfo {author} {\bibfnamefont {I.~A.}\ \bibnamefont {Finneran}}, \bibinfo {author} {\bibfnamefont {L.}~\bibnamefont {Anderegg}}, \bibinfo {author} {\bibfnamefont {B.~L.}\ \bibnamefont {Augenbraun}}, \bibinfo {author} {\bibfnamefont {J.~M.}\ \bibnamefont {Doyle}},\ and\ \bibinfo {author} {\bibfnamefont {J.}~\bibnamefont {Ye}},\ }\bibfield  {title} {\bibinfo {title} {{3D Magneto-Optical Trap of Yttrium Monoxide}},\ }\href {https://doi.org/10.1103/PhysRevLett.121.213201} {\bibfield  {journal} {\bibinfo  {journal} {Phys. Rev. Lett.}\ }\textbf {\bibinfo {volume} {121}},\ \bibinfo {pages} {213201} (\bibinfo {year} {2018})}\BibitemShut {NoStop}%
\bibitem [{\citenamefont {Anderegg}\ \emph {et~al.}(2017)\citenamefont {Anderegg}, \citenamefont {Augenbraun}, \citenamefont {Chae}, \citenamefont {Hemmerling}, \citenamefont {Hutzler}, \citenamefont {Ravi}, \citenamefont {Collopy}, \citenamefont {Ye}, \citenamefont {Ketterle},\ and\ \citenamefont {Doyle}}]{Cafradio}%
  \BibitemOpen
  \bibfield  {author} {\bibinfo {author} {\bibfnamefont {L.}~\bibnamefont {Anderegg}}, \bibinfo {author} {\bibfnamefont {B.~L.}\ \bibnamefont {Augenbraun}}, \bibinfo {author} {\bibfnamefont {E.}~\bibnamefont {Chae}}, \bibinfo {author} {\bibfnamefont {B.}~\bibnamefont {Hemmerling}}, \bibinfo {author} {\bibfnamefont {N.~R.}\ \bibnamefont {Hutzler}}, \bibinfo {author} {\bibfnamefont {A.}~\bibnamefont {Ravi}}, \bibinfo {author} {\bibfnamefont {A.}~\bibnamefont {Collopy}}, \bibinfo {author} {\bibfnamefont {J.}~\bibnamefont {Ye}}, \bibinfo {author} {\bibfnamefont {W.}~\bibnamefont {Ketterle}},\ and\ \bibinfo {author} {\bibfnamefont {J.~M.}\ \bibnamefont {Doyle}},\ }\bibfield  {title} {\bibinfo {title} {{Radio Frequency Magneto-Optical Trapping of CaF with High Density}},\ }\href {https://doi.org/10.1103/PhysRevLett.119.103201} {\bibfield  {journal} {\bibinfo  {journal} {Phys. Rev. Lett.}\ }\textbf {\bibinfo {volume} {119}},\ \bibinfo {pages} {103201} (\bibinfo {year} {2017})}\BibitemShut {NoStop}%
\bibitem [{\citenamefont {Vilas}\ \emph {et~al.}(2022)\citenamefont {Vilas}, \citenamefont {Hallas}, \citenamefont {Anderegg}, \citenamefont {Robichaud}, \citenamefont {Winnicki}, \citenamefont {Mitra},\ and\ \citenamefont {Doyle}}]{Vilas_2022}%
  \BibitemOpen
  \bibfield  {author} {\bibinfo {author} {\bibfnamefont {N.~B.}\ \bibnamefont {Vilas}}, \bibinfo {author} {\bibfnamefont {C.}~\bibnamefont {Hallas}}, \bibinfo {author} {\bibfnamefont {L.}~\bibnamefont {Anderegg}}, \bibinfo {author} {\bibfnamefont {P.}~\bibnamefont {Robichaud}}, \bibinfo {author} {\bibfnamefont {A.}~\bibnamefont {Winnicki}}, \bibinfo {author} {\bibfnamefont {D.}~\bibnamefont {Mitra}},\ and\ \bibinfo {author} {\bibfnamefont {J.~M.}\ \bibnamefont {Doyle}},\ }\bibfield  {title} {\bibinfo {title} {{Magneto-optical Trapping and Sub-Doppler Cooling of a Polyatomic Molecule}},\ }\href {https://doi.org/10.1038/s41586-022-04620-5} {\bibfield  {journal} {\bibinfo  {journal} {Nature}\ }\textbf {\bibinfo {volume} {606}},\ \bibinfo {pages} {70–74} (\bibinfo {year} {2022})}\BibitemShut {NoStop}%
\bibitem [{\citenamefont {Ding}\ \emph {et~al.}(2020)\citenamefont {Ding}, \citenamefont {Wu}, \citenamefont {Finneran}, \citenamefont {Burau},\ and\ \citenamefont {Ye}}]{ding2020}%
  \BibitemOpen
  \bibfield  {author} {\bibinfo {author} {\bibfnamefont {S.}~\bibnamefont {Ding}}, \bibinfo {author} {\bibfnamefont {Y.}~\bibnamefont {Wu}}, \bibinfo {author} {\bibfnamefont {I.~A.}\ \bibnamefont {Finneran}}, \bibinfo {author} {\bibfnamefont {J.~J.}\ \bibnamefont {Burau}},\ and\ \bibinfo {author} {\bibfnamefont {J.}~\bibnamefont {Ye}},\ }\bibfield  {title} {\bibinfo {title} {{Sub-Doppler Cooling and Compressed Trapping of YO Molecules at $\ensuremath{\mu}\mathrm{K}$ Temperatures}},\ }\href {https://doi.org/10.1103/PhysRevX.10.021049} {\bibfield  {journal} {\bibinfo  {journal} {Phys. Rev. X}\ }\textbf {\bibinfo {volume} {10}},\ \bibinfo {pages} {021049} (\bibinfo {year} {2020})}\BibitemShut {NoStop}%
\bibitem [{\citenamefont {Cheuk}\ \emph {et~al.}(2018)\citenamefont {Cheuk}, \citenamefont {Anderegg}, \citenamefont {Augenbraun}, \citenamefont {Bao}, \citenamefont {Burchesky}, \citenamefont {Ketterle},\ and\ \citenamefont {Doyle}}]{caflambda}%
  \BibitemOpen
  \bibfield  {author} {\bibinfo {author} {\bibfnamefont {L.~W.}\ \bibnamefont {Cheuk}}, \bibinfo {author} {\bibfnamefont {L.}~\bibnamefont {Anderegg}}, \bibinfo {author} {\bibfnamefont {B.~L.}\ \bibnamefont {Augenbraun}}, \bibinfo {author} {\bibfnamefont {Y.}~\bibnamefont {Bao}}, \bibinfo {author} {\bibfnamefont {S.}~\bibnamefont {Burchesky}}, \bibinfo {author} {\bibfnamefont {W.}~\bibnamefont {Ketterle}},\ and\ \bibinfo {author} {\bibfnamefont {J.~M.}\ \bibnamefont {Doyle}},\ }\bibfield  {title} {\bibinfo {title} {{$\mathrm{\ensuremath{\Lambda}}$-Enhanced Imaging of Molecules in an Optical Trap}},\ }\href {https://doi.org/10.1103/PhysRevLett.121.083201} {\bibfield  {journal} {\bibinfo  {journal} {Phys. Rev. Lett.}\ }\textbf {\bibinfo {volume} {121}},\ \bibinfo {pages} {083201} (\bibinfo {year} {2018})}\BibitemShut {NoStop}%
\bibitem [{\citenamefont {Caldwell}\ \emph {et~al.}(2019)\citenamefont {Caldwell}, \citenamefont {Devlin}, \citenamefont {Williams}, \citenamefont {Fitch}, \citenamefont {Hinds}, \citenamefont {Sauer},\ and\ \citenamefont {Tarbutt}}]{deep}%
  \BibitemOpen
  \bibfield  {author} {\bibinfo {author} {\bibfnamefont {L.}~\bibnamefont {Caldwell}}, \bibinfo {author} {\bibfnamefont {J.~A.}\ \bibnamefont {Devlin}}, \bibinfo {author} {\bibfnamefont {H.~J.}\ \bibnamefont {Williams}}, \bibinfo {author} {\bibfnamefont {N.~J.}\ \bibnamefont {Fitch}}, \bibinfo {author} {\bibfnamefont {E.~A.}\ \bibnamefont {Hinds}}, \bibinfo {author} {\bibfnamefont {B.~E.}\ \bibnamefont {Sauer}},\ and\ \bibinfo {author} {\bibfnamefont {M.~R.}\ \bibnamefont {Tarbutt}},\ }\bibfield  {title} {\bibinfo {title} {{Deep Laser Cooling and Efficient Magnetic Compression of Molecules}},\ }\href {https://doi.org/10.1103/PhysRevLett.123.033202} {\bibfield  {journal} {\bibinfo  {journal} {Phys. Rev. Lett.}\ }\textbf {\bibinfo {volume} {123}},\ \bibinfo {pages} {033202} (\bibinfo {year} {2019})}\BibitemShut {NoStop}%
\bibitem [{\citenamefont {McCarron}\ \emph {et~al.}(2018)\citenamefont {McCarron}, \citenamefont {Steinecker}, \citenamefont {Zhu},\ and\ \citenamefont {DeMille}}]{magnetic}%
  \BibitemOpen
  \bibfield  {author} {\bibinfo {author} {\bibfnamefont {D.~J.}\ \bibnamefont {McCarron}}, \bibinfo {author} {\bibfnamefont {M.~H.}\ \bibnamefont {Steinecker}}, \bibinfo {author} {\bibfnamefont {Y.}~\bibnamefont {Zhu}},\ and\ \bibinfo {author} {\bibfnamefont {D.}~\bibnamefont {DeMille}},\ }\bibfield  {title} {\bibinfo {title} {{Magnetic Trapping of an Ultracold Gas of Polar Molecules}},\ }\href {https://doi.org/10.1103/PhysRevLett.121.013202} {\bibfield  {journal} {\bibinfo  {journal} {Phys. Rev. Lett.}\ }\textbf {\bibinfo {volume} {121}},\ \bibinfo {pages} {013202} (\bibinfo {year} {2018})}\BibitemShut {NoStop}%
\bibitem [{\citenamefont {Roussy}\ \emph {et~al.}(2023)\citenamefont {Roussy}, \citenamefont {Caldwell}, \citenamefont {Wright}, \citenamefont {Cairncross}, \citenamefont {Shagam}, \citenamefont {Ng}, \citenamefont {Schlossberger}, \citenamefont {Park}, \citenamefont {Wang}, \citenamefont {Ye},\ and\ \citenamefont {Cornell}}]{Roussy2023}%
  \BibitemOpen
  \bibfield  {author} {\bibinfo {author} {\bibfnamefont {T.~S.}\ \bibnamefont {Roussy}}, \bibinfo {author} {\bibfnamefont {L.}~\bibnamefont {Caldwell}}, \bibinfo {author} {\bibfnamefont {T.}~\bibnamefont {Wright}}, \bibinfo {author} {\bibfnamefont {W.~B.}\ \bibnamefont {Cairncross}}, \bibinfo {author} {\bibfnamefont {Y.}~\bibnamefont {Shagam}}, \bibinfo {author} {\bibfnamefont {K.~B.}\ \bibnamefont {Ng}}, \bibinfo {author} {\bibfnamefont {N.}~\bibnamefont {Schlossberger}}, \bibinfo {author} {\bibfnamefont {S.~Y.}\ \bibnamefont {Park}}, \bibinfo {author} {\bibfnamefont {A.}~\bibnamefont {Wang}}, \bibinfo {author} {\bibfnamefont {J.}~\bibnamefont {Ye}},\ and\ \bibinfo {author} {\bibfnamefont {E.~A.}\ \bibnamefont {Cornell}},\ }\bibfield  {title} {\bibinfo {title} {{An Improved Bound on the Electron’s Electric Dipole Moment}},\ }\href {https://doi.org/10.1126/science.adg4084} {\bibfield  {journal} {\bibinfo  {journal} {Science}\ }\textbf {\bibinfo {volume} {381}},\ \bibinfo {pages} {46} (\bibinfo {year}
  {2023})}\BibitemShut {NoStop}%
\bibitem [{\citenamefont {ACME-Collaboration}(2018)}]{ACME2018}%
  \BibitemOpen
  \bibfield  {author} {\bibinfo {author} {\bibnamefont {ACME-Collaboration}},\ }\bibfield  {title} {\bibinfo {title} {{Improved Limit on the Electric Dipole Moment of the Electron}},\ }\href {https://doi.org/10.1038/s41586-018-0599-8} {\bibfield  {journal} {\bibinfo  {journal} {Nature}\ }\textbf {\bibinfo {volume} {562}},\ \bibinfo {pages} {355–360} (\bibinfo {year} {2018})}\BibitemShut {NoStop}%
\bibitem [{\citenamefont {Liu}\ \emph {et~al.}(2021)\citenamefont {Liu}, \citenamefont {Hu}, \citenamefont {Nichols}, \citenamefont {Yang}, \citenamefont {Xie}, \citenamefont {Guo},\ and\ \citenamefont {Ni}}]{Liu_2021}%
  \BibitemOpen
  \bibfield  {author} {\bibinfo {author} {\bibfnamefont {Y.}~\bibnamefont {Liu}}, \bibinfo {author} {\bibfnamefont {M.-G.}\ \bibnamefont {Hu}}, \bibinfo {author} {\bibfnamefont {M.~A.}\ \bibnamefont {Nichols}}, \bibinfo {author} {\bibfnamefont {D.}~\bibnamefont {Yang}}, \bibinfo {author} {\bibfnamefont {D.}~\bibnamefont {Xie}}, \bibinfo {author} {\bibfnamefont {H.}~\bibnamefont {Guo}},\ and\ \bibinfo {author} {\bibfnamefont {K.-K.}\ \bibnamefont {Ni}},\ }\bibfield  {title} {\bibinfo {title} {{Precision Test of Statistical Dynamics with State-to-state Ultracold Chemistry}},\ }\href {https://doi.org/10.1038/s41586-021-03459-6} {\bibfield  {journal} {\bibinfo  {journal} {Nature}\ }\textbf {\bibinfo {volume} {593}},\ \bibinfo {pages} {379–384} (\bibinfo {year} {2021})}\BibitemShut {NoStop}%
\bibitem [{\citenamefont {Son}\ \emph {et~al.}(2022)\citenamefont {Son}, \citenamefont {Park}, \citenamefont {Lu}, \citenamefont {Jamison}, \citenamefont {Karman},\ and\ \citenamefont {Ketterle}}]{Son_2022}%
  \BibitemOpen
  \bibfield  {author} {\bibinfo {author} {\bibfnamefont {H.}~\bibnamefont {Son}}, \bibinfo {author} {\bibfnamefont {J.~J.}\ \bibnamefont {Park}}, \bibinfo {author} {\bibfnamefont {Y.-K.}\ \bibnamefont {Lu}}, \bibinfo {author} {\bibfnamefont {A.~O.}\ \bibnamefont {Jamison}}, \bibinfo {author} {\bibfnamefont {T.}~\bibnamefont {Karman}},\ and\ \bibinfo {author} {\bibfnamefont {W.}~\bibnamefont {Ketterle}},\ }\bibfield  {title} {\bibinfo {title} {{Control of Reactive Collisions by Quantum Interference}},\ }\href {https://doi.org/10.1126/science.abl7257} {\bibfield  {journal} {\bibinfo  {journal} {Science}\ }\textbf {\bibinfo {volume} {375}},\ \bibinfo {pages} {1006–1010} (\bibinfo {year} {2022})}\BibitemShut {NoStop}%
\bibitem [{\citenamefont {Chen}\ \emph {et~al.}(2024)\citenamefont {Chen}, \citenamefont {Biswas}, \citenamefont {Eppelt}, \citenamefont {Schindewolf}, \citenamefont {Deng}, \citenamefont {Shi}, \citenamefont {Yi}, \citenamefont {Hilker}, \citenamefont {Bloch},\ and\ \citenamefont {Luo}}]{Chen2024}%
  \BibitemOpen
  \bibfield  {author} {\bibinfo {author} {\bibfnamefont {X.-Y.}\ \bibnamefont {Chen}}, \bibinfo {author} {\bibfnamefont {S.}~\bibnamefont {Biswas}}, \bibinfo {author} {\bibfnamefont {S.}~\bibnamefont {Eppelt}}, \bibinfo {author} {\bibfnamefont {A.}~\bibnamefont {Schindewolf}}, \bibinfo {author} {\bibfnamefont {F.}~\bibnamefont {Deng}}, \bibinfo {author} {\bibfnamefont {T.}~\bibnamefont {Shi}}, \bibinfo {author} {\bibfnamefont {S.}~\bibnamefont {Yi}}, \bibinfo {author} {\bibfnamefont {T.~A.}\ \bibnamefont {Hilker}}, \bibinfo {author} {\bibfnamefont {I.}~\bibnamefont {Bloch}},\ and\ \bibinfo {author} {\bibfnamefont {X.-Y.}\ \bibnamefont {Luo}},\ }\bibfield  {title} {\bibinfo {title} {{Ultracold Field-linked Tetratomic Molecules}},\ }\bibfield  {journal} {\bibinfo  {journal} {Nature}\ }\href {https://doi.org/10.1038/s41586-023-06986-6} {10.1038/s41586-023-06986-6} (\bibinfo {year} {2024})\BibitemShut {NoStop}%
\bibitem [{\citenamefont {Anderson}\ \emph {et~al.}(1995)\citenamefont {Anderson}, \citenamefont {Ensher}, \citenamefont {Matthews}, \citenamefont {Wieman},\ and\ \citenamefont {Cornell}}]{Bec}%
  \BibitemOpen
  \bibfield  {author} {\bibinfo {author} {\bibfnamefont {M.~H.}\ \bibnamefont {Anderson}}, \bibinfo {author} {\bibfnamefont {J.~R.}\ \bibnamefont {Ensher}}, \bibinfo {author} {\bibfnamefont {M.~R.}\ \bibnamefont {Matthews}}, \bibinfo {author} {\bibfnamefont {C.~E.}\ \bibnamefont {Wieman}},\ and\ \bibinfo {author} {\bibfnamefont {E.~A.}\ \bibnamefont {Cornell}},\ }\bibfield  {title} {\bibinfo {title} {{Observation of Bose-Einstein Condensation in a Dilute Atomic Vapor}},\ }\href {https://doi.org/10.1126/science.269.5221.198} {\bibfield  {journal} {\bibinfo  {journal} {Science}\ }\textbf {\bibinfo {volume} {269}},\ \bibinfo {pages} {198} (\bibinfo {year} {1995})}\BibitemShut {NoStop}%
\bibitem [{\citenamefont {Davis}\ \emph {et~al.}(1995)\citenamefont {Davis}, \citenamefont {Mewes}, \citenamefont {Andrews}, \citenamefont {van Druten}, \citenamefont {Durfee}, \citenamefont {Kurn},\ and\ \citenamefont {Ketterle}}]{ketterle}%
  \BibitemOpen
  \bibfield  {author} {\bibinfo {author} {\bibfnamefont {K.~B.}\ \bibnamefont {Davis}}, \bibinfo {author} {\bibfnamefont {M.~O.}\ \bibnamefont {Mewes}}, \bibinfo {author} {\bibfnamefont {M.~R.}\ \bibnamefont {Andrews}}, \bibinfo {author} {\bibfnamefont {N.~J.}\ \bibnamefont {van Druten}}, \bibinfo {author} {\bibfnamefont {D.~S.}\ \bibnamefont {Durfee}}, \bibinfo {author} {\bibfnamefont {D.~M.}\ \bibnamefont {Kurn}},\ and\ \bibinfo {author} {\bibfnamefont {W.}~\bibnamefont {Ketterle}},\ }\bibfield  {title} {\bibinfo {title} {{Bose-Einstein Condensation in a Gas of Sodium Atoms}},\ }\href {https://doi.org/10.1103/PhysRevLett.75.3969} {\bibfield  {journal} {\bibinfo  {journal} {Phys. Rev. Lett.}\ }\textbf {\bibinfo {volume} {75}},\ \bibinfo {pages} {3969} (\bibinfo {year} {1995})}\BibitemShut {NoStop}%
\bibitem [{\citenamefont {Wu}\ \emph {et~al.}(2021)\citenamefont {Wu}, \citenamefont {Burau}, \citenamefont {Mehling}, \citenamefont {Ye},\ and\ \citenamefont {Ding}}]{Wu2021}%
  \BibitemOpen
  \bibfield  {author} {\bibinfo {author} {\bibfnamefont {Y.}~\bibnamefont {Wu}}, \bibinfo {author} {\bibfnamefont {J.~J.}\ \bibnamefont {Burau}}, \bibinfo {author} {\bibfnamefont {K.}~\bibnamefont {Mehling}}, \bibinfo {author} {\bibfnamefont {J.}~\bibnamefont {Ye}},\ and\ \bibinfo {author} {\bibfnamefont {S.}~\bibnamefont {Ding}},\ }\bibfield  {title} {\bibinfo {title} {{High Phase-Space Density of Laser-Cooled Molecules in an Optical Lattice}},\ }\href {https://doi.org/10.1103/PhysRevLett.127.263201} {\bibfield  {journal} {\bibinfo  {journal} {Phys. Rev. Lett.}\ }\textbf {\bibinfo {volume} {127}},\ \bibinfo {pages} {263201} (\bibinfo {year} {2021})}\BibitemShut {NoStop}%
\bibitem [{\citenamefont {Burau}\ \emph {et~al.}(2023)\citenamefont {Burau}, \citenamefont {Aggarwal}, \citenamefont {Mehling},\ and\ \citenamefont {Ye}}]{Burau2023}%
  \BibitemOpen
  \bibfield  {author} {\bibinfo {author} {\bibfnamefont {J.~J.}\ \bibnamefont {Burau}}, \bibinfo {author} {\bibfnamefont {P.}~\bibnamefont {Aggarwal}}, \bibinfo {author} {\bibfnamefont {K.}~\bibnamefont {Mehling}},\ and\ \bibinfo {author} {\bibfnamefont {J.}~\bibnamefont {Ye}},\ }\bibfield  {title} {\bibinfo {title} {{Blue-Detuned Magneto-optical Trap of Molecules}},\ }\href {https://doi.org/10.1103/PhysRevLett.130.193401} {\bibfield  {journal} {\bibinfo  {journal} {Phys. Rev. Lett.}\ }\textbf {\bibinfo {volume} {130}},\ \bibinfo {pages} {193401} (\bibinfo {year} {2023})}\BibitemShut {NoStop}%
\bibitem [{\citenamefont {Jorapur}\ \emph {et~al.}(2024)\citenamefont {Jorapur}, \citenamefont {Langin}, \citenamefont {Wang}, \citenamefont {Zheng},\ and\ \citenamefont {DeMille}}]{jorapur2023high}%
  \BibitemOpen
  \bibfield  {author} {\bibinfo {author} {\bibfnamefont {V.}~\bibnamefont {Jorapur}}, \bibinfo {author} {\bibfnamefont {T.~K.}\ \bibnamefont {Langin}}, \bibinfo {author} {\bibfnamefont {Q.}~\bibnamefont {Wang}}, \bibinfo {author} {\bibfnamefont {G.}~\bibnamefont {Zheng}},\ and\ \bibinfo {author} {\bibfnamefont {D.}~\bibnamefont {DeMille}},\ }\bibfield  {title} {\bibinfo {title} {{High Density Loading and Collisional Loss of Laser-Cooled Molecules in an Optical Trap}},\ }\href {https://doi.org/10.1103/PhysRevLett.132.163403} {\bibfield  {journal} {\bibinfo  {journal} {Phys. Rev. Lett.}\ }\textbf {\bibinfo {volume} {132}},\ \bibinfo {pages} {163403} (\bibinfo {year} {2024})}\BibitemShut {NoStop}%
\bibitem [{\citenamefont {Li}\ \emph {et~al.}(2023{\natexlab{b}})\citenamefont {Li}, \citenamefont {Holland}, \citenamefont {Lu},\ and\ \citenamefont {Cheuk}}]{li2023bluedetuned}%
  \BibitemOpen
  \bibfield  {author} {\bibinfo {author} {\bibfnamefont {S.~J.}\ \bibnamefont {Li}}, \bibinfo {author} {\bibfnamefont {C.~M.}\ \bibnamefont {Holland}}, \bibinfo {author} {\bibfnamefont {Y.}~\bibnamefont {Lu}},\ and\ \bibinfo {author} {\bibfnamefont {L.~W.}\ \bibnamefont {Cheuk}},\ }\href@noop {} {\bibinfo {title} {{A Blue-Detuned Magneto-Optical Trap of CaF Molecules}}} (\bibinfo {year} {2023}{\natexlab{b}}),\ \Eprint {https://arxiv.org/abs/2311.05447} {arXiv:2311.05447 [physics.atom-ph]} \BibitemShut {NoStop}%
\bibitem [{\citenamefont {Hallas}\ \emph {et~al.}(2024)\citenamefont {Hallas}, \citenamefont {Li}, \citenamefont {Vilas}, \citenamefont {Robichaud}, \citenamefont {Anderegg},\ and\ \citenamefont {Doyle}}]{hallas2024high}%
  \BibitemOpen
  \bibfield  {author} {\bibinfo {author} {\bibfnamefont {C.}~\bibnamefont {Hallas}}, \bibinfo {author} {\bibfnamefont {G.~K.}\ \bibnamefont {Li}}, \bibinfo {author} {\bibfnamefont {N.~B.}\ \bibnamefont {Vilas}}, \bibinfo {author} {\bibfnamefont {P.}~\bibnamefont {Robichaud}}, \bibinfo {author} {\bibfnamefont {L.}~\bibnamefont {Anderegg}},\ and\ \bibinfo {author} {\bibfnamefont {J.~M.}\ \bibnamefont {Doyle}},\ }\href@noop {} {\bibinfo {title} {{High Compression Blue-Detuned Magneto-Optical Trap of Polyatomic Molecules}}} (\bibinfo {year} {2024}),\ \Eprint {https://arxiv.org/abs/2404.03636} {arXiv:2404.03636 [physics.atom-ph]} \BibitemShut {NoStop}%
\bibitem [{\citenamefont {Wang}\ and\ \citenamefont {Quéméner}(2015)}]{Wang_2015electric}%
  \BibitemOpen
  \bibfield  {author} {\bibinfo {author} {\bibfnamefont {G.}~\bibnamefont {Wang}}\ and\ \bibinfo {author} {\bibfnamefont {G.}~\bibnamefont {Quéméner}},\ }\bibfield  {title} {\bibinfo {title} {{Tuning Ultracold Collisions of Excited Rotational Dipolar Molecules}},\ }\href {https://doi.org/10.1088/1367-2630/17/3/035015} {\bibfield  {journal} {\bibinfo  {journal} {New Journal of Physics}\ }\textbf {\bibinfo {volume} {17}},\ \bibinfo {pages} {035015} (\bibinfo {year} {2015})}\BibitemShut {NoStop}%
\bibitem [{\citenamefont {Avdeenkov}\ \emph {et~al.}(2006)\citenamefont {Avdeenkov}, \citenamefont {Kajita},\ and\ \citenamefont {Bohn}}]{electricBohn}%
  \BibitemOpen
  \bibfield  {author} {\bibinfo {author} {\bibfnamefont {A.~V.}\ \bibnamefont {Avdeenkov}}, \bibinfo {author} {\bibfnamefont {M.}~\bibnamefont {Kajita}},\ and\ \bibinfo {author} {\bibfnamefont {J.~L.}\ \bibnamefont {Bohn}},\ }\bibfield  {title} {\bibinfo {title} {{Suppression of Inelastic Collisions of Polar $^{1}\ensuremath{\Sigma}$ State Molecules in an Electrostatic Field}},\ }\href {https://doi.org/10.1103/PhysRevA.73.022707} {\bibfield  {journal} {\bibinfo  {journal} {Phys. Rev. A}\ }\textbf {\bibinfo {volume} {73}},\ \bibinfo {pages} {022707} (\bibinfo {year} {2006})}\BibitemShut {NoStop}%
\bibitem [{\citenamefont {Matsuda}\ \emph {et~al.}(2020)\citenamefont {Matsuda}, \citenamefont {De~Marco}, \citenamefont {Li}, \citenamefont {Tobias}, \citenamefont {Valtolina}, \citenamefont {Quéméner},\ and\ \citenamefont {Ye}}]{Matsuda_2020}%
  \BibitemOpen
  \bibfield  {author} {\bibinfo {author} {\bibfnamefont {K.}~\bibnamefont {Matsuda}}, \bibinfo {author} {\bibfnamefont {L.}~\bibnamefont {De~Marco}}, \bibinfo {author} {\bibfnamefont {J.-R.}\ \bibnamefont {Li}}, \bibinfo {author} {\bibfnamefont {W.~G.}\ \bibnamefont {Tobias}}, \bibinfo {author} {\bibfnamefont {G.}~\bibnamefont {Valtolina}}, \bibinfo {author} {\bibfnamefont {G.}~\bibnamefont {Quéméner}},\ and\ \bibinfo {author} {\bibfnamefont {J.}~\bibnamefont {Ye}},\ }\bibfield  {title} {\bibinfo {title} {{Resonant Collisional Shielding of Reactive Molecules Using Electric Fields}},\ }\href {https://doi.org/10.1126/science.abe7370} {\bibfield  {journal} {\bibinfo  {journal} {Science}\ }\textbf {\bibinfo {volume} {370}},\ \bibinfo {pages} {1324–1327} (\bibinfo {year} {2020})}\BibitemShut {NoStop}%
\bibitem [{\citenamefont {Mukherjee}\ \emph {et~al.}(2023)\citenamefont {Mukherjee}, \citenamefont {Frye}, \citenamefont {Le~Sueur}, \citenamefont {Tarbutt},\ and\ \citenamefont {Hutson}}]{cafelectric}%
  \BibitemOpen
  \bibfield  {author} {\bibinfo {author} {\bibfnamefont {B.}~\bibnamefont {Mukherjee}}, \bibinfo {author} {\bibfnamefont {M.~D.}\ \bibnamefont {Frye}}, \bibinfo {author} {\bibfnamefont {C.~R.}\ \bibnamefont {Le~Sueur}}, \bibinfo {author} {\bibfnamefont {M.~R.}\ \bibnamefont {Tarbutt}},\ and\ \bibinfo {author} {\bibfnamefont {J.~M.}\ \bibnamefont {Hutson}},\ }\bibfield  {title} {\bibinfo {title} {{Shielding Collisions of Ultracold CaF Molecules with Static Electric Fields}},\ }\href {https://doi.org/10.1103/PhysRevResearch.5.033097} {\bibfield  {journal} {\bibinfo  {journal} {Phys. Rev. Res.}\ }\textbf {\bibinfo {volume} {5}},\ \bibinfo {pages} {033097} (\bibinfo {year} {2023})}\BibitemShut {NoStop}%
\bibitem [{\citenamefont {Karman}\ and\ \citenamefont {Hutson}(2018)}]{tjis}%
  \BibitemOpen
  \bibfield  {author} {\bibinfo {author} {\bibfnamefont {T.}~\bibnamefont {Karman}}\ and\ \bibinfo {author} {\bibfnamefont {J.~M.}\ \bibnamefont {Hutson}},\ }\bibfield  {title} {\bibinfo {title} {{Microwave Shielding of Ultracold Polar Molecules}},\ }\href {https://doi.org/10.1103/PhysRevLett.121.163401} {\bibfield  {journal} {\bibinfo  {journal} {Phys. Rev. Lett.}\ }\textbf {\bibinfo {volume} {121}},\ \bibinfo {pages} {163401} (\bibinfo {year} {2018})}\BibitemShut {NoStop}%
\bibitem [{\citenamefont {Anderegg}\ \emph {et~al.}(2021)\citenamefont {Anderegg}, \citenamefont {Burchesky}, \citenamefont {Bao}, \citenamefont {Yu}, \citenamefont {Karman}, \citenamefont {Chae}, \citenamefont {Ni}, \citenamefont {Ketterle},\ and\ \citenamefont {Doyle}}]{Anderegg2021}%
  \BibitemOpen
  \bibfield  {author} {\bibinfo {author} {\bibfnamefont {L.}~\bibnamefont {Anderegg}}, \bibinfo {author} {\bibfnamefont {S.}~\bibnamefont {Burchesky}}, \bibinfo {author} {\bibfnamefont {Y.}~\bibnamefont {Bao}}, \bibinfo {author} {\bibfnamefont {S.~S.}\ \bibnamefont {Yu}}, \bibinfo {author} {\bibfnamefont {T.}~\bibnamefont {Karman}}, \bibinfo {author} {\bibfnamefont {E.}~\bibnamefont {Chae}}, \bibinfo {author} {\bibfnamefont {K.-K.}\ \bibnamefont {Ni}}, \bibinfo {author} {\bibfnamefont {W.}~\bibnamefont {Ketterle}},\ and\ \bibinfo {author} {\bibfnamefont {J.~M.}\ \bibnamefont {Doyle}},\ }\bibfield  {title} {\bibinfo {title} {{Observation of Microwave Shielding of Ultracold Molecules}},\ }\href {https://doi.org/10.1126/science.abg9502} {\bibfield  {journal} {\bibinfo  {journal} {Science}\ }\textbf {\bibinfo {volume} {373}},\ \bibinfo {pages} {779} (\bibinfo {year} {2021})}\BibitemShut {NoStop}%
\bibitem [{\citenamefont {Lin}\ \emph {et~al.}(2023)\citenamefont {Lin}, \citenamefont {Chen}, \citenamefont {Jin}, \citenamefont {Shi}, \citenamefont {Deng}, \citenamefont {Zhang}, \citenamefont {Qu\'em\'ener}, \citenamefont {Shi}, \citenamefont {Yi},\ and\ \citenamefont {Wang}}]{Hongkong2023}%
  \BibitemOpen
  \bibfield  {author} {\bibinfo {author} {\bibfnamefont {J.}~\bibnamefont {Lin}}, \bibinfo {author} {\bibfnamefont {G.}~\bibnamefont {Chen}}, \bibinfo {author} {\bibfnamefont {M.}~\bibnamefont {Jin}}, \bibinfo {author} {\bibfnamefont {Z.}~\bibnamefont {Shi}}, \bibinfo {author} {\bibfnamefont {F.}~\bibnamefont {Deng}}, \bibinfo {author} {\bibfnamefont {W.}~\bibnamefont {Zhang}}, \bibinfo {author} {\bibfnamefont {G.}~\bibnamefont {Qu\'em\'ener}}, \bibinfo {author} {\bibfnamefont {T.}~\bibnamefont {Shi}}, \bibinfo {author} {\bibfnamefont {S.}~\bibnamefont {Yi}},\ and\ \bibinfo {author} {\bibfnamefont {D.}~\bibnamefont {Wang}},\ }\bibfield  {title} {\bibinfo {title} {{Microwave Shielding of Bosonic {NaRb} Molecules}},\ }\href {https://doi.org/10.1103/PhysRevX.13.031032} {\bibfield  {journal} {\bibinfo  {journal} {Phys. Rev. X}\ }\textbf {\bibinfo {volume} {13}},\ \bibinfo {pages} {031032} (\bibinfo {year} {2023})}\BibitemShut {NoStop}%
\bibitem [{\citenamefont {Li}\ \emph {et~al.}(2021)\citenamefont {Li}, \citenamefont {Tobias}, \citenamefont {Matsuda}, \citenamefont {Miller}, \citenamefont {Valtolina}, \citenamefont {De~Marco}, \citenamefont {Wang}, \citenamefont {Lassabliére}, \citenamefont {Quéméner}, \citenamefont {Bohn},\ and\ \citenamefont {Ye}}]{Li2021}%
  \BibitemOpen
  \bibfield  {author} {\bibinfo {author} {\bibfnamefont {J.-R.}\ \bibnamefont {Li}}, \bibinfo {author} {\bibfnamefont {W.~G.}\ \bibnamefont {Tobias}}, \bibinfo {author} {\bibfnamefont {K.}~\bibnamefont {Matsuda}}, \bibinfo {author} {\bibfnamefont {C.}~\bibnamefont {Miller}}, \bibinfo {author} {\bibfnamefont {G.}~\bibnamefont {Valtolina}}, \bibinfo {author} {\bibfnamefont {L.}~\bibnamefont {De~Marco}}, \bibinfo {author} {\bibfnamefont {R.~R.}\ \bibnamefont {Wang}}, \bibinfo {author} {\bibfnamefont {L.}~\bibnamefont {Lassabliére}}, \bibinfo {author} {\bibfnamefont {G.}~\bibnamefont {Quéméner}}, \bibinfo {author} {\bibfnamefont {J.~L.}\ \bibnamefont {Bohn}},\ and\ \bibinfo {author} {\bibfnamefont {J.}~\bibnamefont {Ye}},\ }\bibfield  {title} {\bibinfo {title} {{Tuning of Dipolar Interactions and Evaporative Cooling in a Three-dimensional Molecular Quantum Gas}},\ }\href@noop {} {\bibfield  {journal} {\bibinfo  {journal} {Nature Phys.}\ }\textbf {\bibinfo {volume} {17}},\ \bibinfo {pages} {1144 – 1148}
  (\bibinfo {year} {2021})}\BibitemShut {NoStop}%
\bibitem [{Note1()}]{Note1}%
  \BibitemOpen
  \bibinfo {note} {Note that some conventions \cite {chengchin} incorporate a factor of 2 within the definition of the cross-section (and so rate coefficient) but not in the rate equation. However, here we follow the convention of Huo and Green \cite {Huo:1996} which places this factor in the rate equation Eq.(\ref {eq:indistingusihablebegin}) and is usually used in molecular scattering as it is clearer for deriving relations with multiple occupied states, and for imposing correct symmetrization.}\BibitemShut {Stop}%
\bibitem [{\citenamefont {Ackermann}\ \emph {et~al.}(1964)\citenamefont {Ackermann}, \citenamefont {Rauh},\ and\ \citenamefont {Thorn}}]{ackermann1964}%
  \BibitemOpen
  \bibfield  {author} {\bibinfo {author} {\bibfnamefont {R.~J.}\ \bibnamefont {Ackermann}}, \bibinfo {author} {\bibfnamefont {E.~G.}\ \bibnamefont {Rauh}},\ and\ \bibinfo {author} {\bibfnamefont {R.~J.}\ \bibnamefont {Thorn}},\ }\bibfield  {title} {\bibinfo {title} {{{Thermodynamic Properties of Gaseous Yttrium Monoxide. Correlation of Bonding in Group III Transition‐Metal Monoxides}}},\ }\href {https://doi.org/10.1063/1.1725221} {\bibfield  {journal} {\bibinfo  {journal} {The Journal of Chemical Physics}\ }\textbf {\bibinfo {volume} {40}},\ \bibinfo {pages} {883} (\bibinfo {year} {1964})}\BibitemShut {NoStop}%
\bibitem [{\citenamefont {Ackermann}\ and\ \citenamefont {Rauh}(1974)}]{ackermann1974}%
  \BibitemOpen
  \bibfield  {author} {\bibinfo {author} {\bibfnamefont {R.}~\bibnamefont {Ackermann}}\ and\ \bibinfo {author} {\bibfnamefont {E.}~\bibnamefont {Rauh}},\ }\bibfield  {title} {\bibinfo {title} {{Thermodynamic Properties of {ZrO} (g) and {HfO} (g); a Critical Examination of Isomolecular Oxygen-exchange Reactions}},\ }\href@noop {} {\bibfield  {journal} {\bibinfo  {journal} {The Journal of Chemical Physics}\ }\textbf {\bibinfo {volume} {60}},\ \bibinfo {pages} {2266} (\bibinfo {year} {1974})}\BibitemShut {NoStop}%
\bibitem [{\citenamefont {Brix}\ and\ \citenamefont {Herzberg}(1954)}]{brix1954}%
  \BibitemOpen
  \bibfield  {author} {\bibinfo {author} {\bibfnamefont {P.}~\bibnamefont {Brix}}\ and\ \bibinfo {author} {\bibfnamefont {G.}~\bibnamefont {Herzberg}},\ }\bibfield  {title} {\bibinfo {title} {{Fine Structure of the {S}chumann-{R}unge Bands Near the Convergence Limit and the Dissociation Energy of the Oxygen Molecule}},\ }\href@noop {} {\bibfield  {journal} {\bibinfo  {journal} {Canadian Journal of Physics}\ }\textbf {\bibinfo {volume} {32}},\ \bibinfo {pages} {110} (\bibinfo {year} {1954})}\BibitemShut {NoStop}%
\bibitem [{\citenamefont {Fang}\ \emph {et~al.}(2000)\citenamefont {Fang}, \citenamefont {Chen}, \citenamefont {Shen}, \citenamefont {Liu}, \citenamefont {Lindsay},\ and\ \citenamefont {Lombardi}}]{fang2000}%
  \BibitemOpen
  \bibfield  {author} {\bibinfo {author} {\bibfnamefont {L.}~\bibnamefont {Fang}}, \bibinfo {author} {\bibfnamefont {X.}~\bibnamefont {Chen}}, \bibinfo {author} {\bibfnamefont {X.}~\bibnamefont {Shen}}, \bibinfo {author} {\bibfnamefont {Y.}~\bibnamefont {Liu}}, \bibinfo {author} {\bibfnamefont {D.}~\bibnamefont {Lindsay}},\ and\ \bibinfo {author} {\bibfnamefont {J.~R.}\ \bibnamefont {Lombardi}},\ }\bibfield  {title} {\bibinfo {title} {{Spectroscopy of Yttrium Dimers in Argon Matrices}},\ }\href@noop {} {\bibfield  {journal} {\bibinfo  {journal} {Low Temperature Physics}\ }\textbf {\bibinfo {volume} {26}},\ \bibinfo {pages} {752} (\bibinfo {year} {2000})}\BibitemShut {NoStop}%
\bibitem [{\citenamefont {Mayle}\ \emph {et~al.}(2012)\citenamefont {Mayle}, \citenamefont {Ruzic},\ and\ \citenamefont {Bohn}}]{mayle2012}%
  \BibitemOpen
  \bibfield  {author} {\bibinfo {author} {\bibfnamefont {M.}~\bibnamefont {Mayle}}, \bibinfo {author} {\bibfnamefont {B.~P.}\ \bibnamefont {Ruzic}},\ and\ \bibinfo {author} {\bibfnamefont {J.~L.}\ \bibnamefont {Bohn}},\ }\bibfield  {title} {\bibinfo {title} {{Statistical Aspects of Ultracold Resonant Scattering}},\ }\href {https://doi.org/10.1103/PhysRevA.85.062712} {\bibfield  {journal} {\bibinfo  {journal} {Phys. Rev. A}\ }\textbf {\bibinfo {volume} {85}},\ \bibinfo {pages} {062712} (\bibinfo {year} {2012})}\BibitemShut {NoStop}%
\bibitem [{\citenamefont {Mayle}\ \emph {et~al.}(2013)\citenamefont {Mayle}, \citenamefont {Qu\'em\'ener}, \citenamefont {Ruzic},\ and\ \citenamefont {Bohn}}]{mayle2013}%
  \BibitemOpen
  \bibfield  {author} {\bibinfo {author} {\bibfnamefont {M.}~\bibnamefont {Mayle}}, \bibinfo {author} {\bibfnamefont {G.}~\bibnamefont {Qu\'em\'ener}}, \bibinfo {author} {\bibfnamefont {B.~P.}\ \bibnamefont {Ruzic}},\ and\ \bibinfo {author} {\bibfnamefont {J.~L.}\ \bibnamefont {Bohn}},\ }\bibfield  {title} {\bibinfo {title} {{Scattering of Ultracold Molecules in the Highly Resonant Regime}},\ }\href {https://doi.org/10.1103/PhysRevA.87.012709} {\bibfield  {journal} {\bibinfo  {journal} {Phys. Rev. A}\ }\textbf {\bibinfo {volume} {87}},\ \bibinfo {pages} {012709} (\bibinfo {year} {2013})}\BibitemShut {NoStop}%
\bibitem [{\citenamefont {Christianen}\ \emph {et~al.}(2019{\natexlab{a}})\citenamefont {Christianen}, \citenamefont {Karman},\ and\ \citenamefont {Groenenboom}}]{Christianen:density:2019}%
  \BibitemOpen
  \bibfield  {author} {\bibinfo {author} {\bibfnamefont {A.}~\bibnamefont {Christianen}}, \bibinfo {author} {\bibfnamefont {T.}~\bibnamefont {Karman}},\ and\ \bibinfo {author} {\bibfnamefont {G.~C.}\ \bibnamefont {Groenenboom}},\ }\bibfield  {title} {\bibinfo {title} {{A {Quasiclassical} Method for Calculating the Density of States of Ultracold Collision Complexes}},\ }\href@noop {} {\bibfield  {journal} {\bibinfo  {journal} {Phys. Rev. A}\ }\textbf {\bibinfo {volume} {100}},\ \bibinfo {pages} {032708} (\bibinfo {year} {2019}{\natexlab{a}})}\BibitemShut {NoStop}%
\bibitem [{\citenamefont {Christianen}\ \emph {et~al.}(2019{\natexlab{b}})\citenamefont {Christianen}, \citenamefont {Zwierlein}, \citenamefont {Groenenboom},\ and\ \citenamefont {Karman}}]{Christianen:laser:2019}%
  \BibitemOpen
  \bibfield  {author} {\bibinfo {author} {\bibfnamefont {A.}~\bibnamefont {Christianen}}, \bibinfo {author} {\bibfnamefont {M.~W.}\ \bibnamefont {Zwierlein}}, \bibinfo {author} {\bibfnamefont {G.~C.}\ \bibnamefont {Groenenboom}},\ and\ \bibinfo {author} {\bibfnamefont {T.}~\bibnamefont {Karman}},\ }\bibfield  {title} {\bibinfo {title} {{Photoinduced Two-Body Loss of Ultracold Molecules}},\ }\href@noop {} {\bibfield  {journal} {\bibinfo  {journal} {Phys. Rev. Lett.}\ }\textbf {\bibinfo {volume} {123}},\ \bibinfo {pages} {123402} (\bibinfo {year} {2019}{\natexlab{b}})}\BibitemShut {NoStop}%
\bibitem [{\citenamefont {Idziaszek}\ and\ \citenamefont {Julienne}(2010)}]{Idziaszek:PRL:2010}%
  \BibitemOpen
  \bibfield  {author} {\bibinfo {author} {\bibfnamefont {Z.}~\bibnamefont {Idziaszek}}\ and\ \bibinfo {author} {\bibfnamefont {P.~S.}\ \bibnamefont {Julienne}},\ }\bibfield  {title} {\bibinfo {title} {{Universal {R}ate {C}onstants for {R}eactive {C}ollisions of {U}ltracold {M}olecules}},\ }\href@noop {} {\bibfield  {journal} {\bibinfo  {journal} {Phys. Rev. Lett.}\ ,\ \bibinfo {pages} {113202}} (\bibinfo {year} {2010})}\BibitemShut {NoStop}%
\bibitem [{\citenamefont {Frye}\ \emph {et~al.}(2015)\citenamefont {Frye}, \citenamefont {Julienne},\ and\ \citenamefont {Hutson}}]{Frye:2015}%
  \BibitemOpen
  \bibfield  {author} {\bibinfo {author} {\bibfnamefont {M.~D.}\ \bibnamefont {Frye}}, \bibinfo {author} {\bibfnamefont {P.~S.}\ \bibnamefont {Julienne}},\ and\ \bibinfo {author} {\bibfnamefont {J.~M.}\ \bibnamefont {Hutson}},\ }\bibfield  {title} {\bibinfo {title} {{Cold atomic and Molecular Collisions: Approaching the Universal Loss Regime}},\ }\href@noop {} {\bibfield  {journal} {\bibinfo  {journal} {New J. Phys.}\ }\textbf {\bibinfo {volume} {17}},\ \bibinfo {pages} {045019} (\bibinfo {year} {2015})}\BibitemShut {NoStop}%
\bibitem [{\citenamefont {Gao}(2008)}]{Gao:2008}%
  \BibitemOpen
  \bibfield  {author} {\bibinfo {author} {\bibfnamefont {B.}~\bibnamefont {Gao}},\ }\bibfield  {title} {\bibinfo {title} {{General Form of the Quantum-defect Theory for $-1/r^\alpha$ Type of Potentials with $\alpha>2$}},\ }\href@noop {} {\bibfield  {journal} {\bibinfo  {journal} {Phys. Rev. A}\ }\textbf {\bibinfo {volume} {78}},\ \bibinfo {pages} {012702} (\bibinfo {year} {2008})}\BibitemShut {NoStop}%
\bibitem [{\citenamefont {Ospelkaus}\ \emph {et~al.}(2010)\citenamefont {Ospelkaus}, \citenamefont {Ni}, \citenamefont {Wang}, \citenamefont {{de Miranda}}, \citenamefont {Neyenhuis}, \citenamefont {Qu\'{e}m\'{e}ner}, \citenamefont {Julienne}, \citenamefont {Bohn}, \citenamefont {Jin},\ and\ \citenamefont {Ye}}]{Ospelkaus:react:2010}%
  \BibitemOpen
  \bibfield  {author} {\bibinfo {author} {\bibfnamefont {S.}~\bibnamefont {Ospelkaus}}, \bibinfo {author} {\bibfnamefont {K.-K.}\ \bibnamefont {Ni}}, \bibinfo {author} {\bibfnamefont {D.}~\bibnamefont {Wang}}, \bibinfo {author} {\bibfnamefont {M.~H.~G.}\ \bibnamefont {{de Miranda}}}, \bibinfo {author} {\bibfnamefont {B.}~\bibnamefont {Neyenhuis}}, \bibinfo {author} {\bibfnamefont {G.}~\bibnamefont {Qu\'{e}m\'{e}ner}}, \bibinfo {author} {\bibfnamefont {P.~S.}\ \bibnamefont {Julienne}}, \bibinfo {author} {\bibfnamefont {J.~L.}\ \bibnamefont {Bohn}}, \bibinfo {author} {\bibfnamefont {D.~S.}\ \bibnamefont {Jin}},\ and\ \bibinfo {author} {\bibfnamefont {J.}~\bibnamefont {Ye}},\ }\bibfield  {title} {\bibinfo {title} {{Quantum-state Controlled Chemical Reactions of Ultracold {KRb} Molecules}},\ }\href@noop {} {\bibfield  {journal} {\bibinfo  {journal} {Science}\ }\textbf {\bibinfo {volume} {327}},\ \bibinfo {pages} {853} (\bibinfo {year} {2010})}\BibitemShut {NoStop}%
\bibitem [{\citenamefont {Gregory}\ \emph {et~al.}(2019)\citenamefont {Gregory}, \citenamefont {Frye}, \citenamefont {Blackmore}, \citenamefont {Bridge}, \citenamefont {Sawant}, \citenamefont {Hutson},\ and\ \citenamefont {Cornish}}]{Gregory_2019}%
  \BibitemOpen
  \bibfield  {author} {\bibinfo {author} {\bibfnamefont {P.~D.}\ \bibnamefont {Gregory}}, \bibinfo {author} {\bibfnamefont {M.~D.}\ \bibnamefont {Frye}}, \bibinfo {author} {\bibfnamefont {J.~A.}\ \bibnamefont {Blackmore}}, \bibinfo {author} {\bibfnamefont {E.~M.}\ \bibnamefont {Bridge}}, \bibinfo {author} {\bibfnamefont {R.}~\bibnamefont {Sawant}}, \bibinfo {author} {\bibfnamefont {J.~M.}\ \bibnamefont {Hutson}},\ and\ \bibinfo {author} {\bibfnamefont {S.~L.}\ \bibnamefont {Cornish}},\ }\bibfield  {title} {\bibinfo {title} {{Sticky Collisions of Ultracold RbCs Molecules}},\ }\href@noop {} {\bibfield  {journal} {\bibinfo  {journal} {Nature Communications}\ }\textbf {\bibinfo {volume} {10}} (\bibinfo {year} {2019})}\BibitemShut {NoStop}%
\bibitem [{\citenamefont {Gao}(1998)}]{Gao:C6:1998}%
  \BibitemOpen
  \bibfield  {author} {\bibinfo {author} {\bibfnamefont {B.}~\bibnamefont {Gao}},\ }\bibfield  {title} {\bibinfo {title} {{Solutions of the {S}chr\"odinger Equation for an Attractive $1/r^6$ Potential}},\ }\href {https://doi.org/10.1103/PhysRevA.58.1728} {\bibfield  {journal} {\bibinfo  {journal} {Phys. Rev. A}\ }\textbf {\bibinfo {volume} {58}},\ \bibinfo {pages} {1728} (\bibinfo {year} {1998})}\BibitemShut {NoStop}%
\bibitem [{\citenamefont {Gao}(2003)}]{Gao:AQDTroutines}%
  \BibitemOpen
  \bibfield  {author} {\bibinfo {author} {\bibfnamefont {B.}~\bibnamefont {Gao}},\ }\bibfield  {title} {\bibinfo {title} {{Routines to Calculate the {AQDT} Parameters for an Attractive $1/r^6$ Potential, {Version 2}}}} (\bibinfo {year} {2003}),\ \bibinfo {note} {{{U}niversity of {T}oledo, {O}hio}}\BibitemShut {NoStop}%
\bibitem [{\citenamefont {Suenram}\ \emph {et~al.}(1990)\citenamefont {Suenram}, \citenamefont {Lovas}, \citenamefont {Fraser},\ and\ \citenamefont {Matsumura}}]{microwave}%
  \BibitemOpen
  \bibfield  {author} {\bibinfo {author} {\bibfnamefont {R.~D.}\ \bibnamefont {Suenram}}, \bibinfo {author} {\bibfnamefont {F.~J.}\ \bibnamefont {Lovas}}, \bibinfo {author} {\bibfnamefont {G.~T.}\ \bibnamefont {Fraser}},\ and\ \bibinfo {author} {\bibfnamefont {K.}~\bibnamefont {Matsumura}},\ }\bibfield  {title} {\bibinfo {title} {{{Pulsed‐nozzle Fourier‐transform Microwave Spectroscopy of Laser‐vaporized Metal Oxides: Rotational Spectra and Electric Dipole Moments of YO, LaO, ZrO, and HfO}}},\ }\href {https://doi.org/10.1063/1.457690} {\bibfield  {journal} {\bibinfo  {journal} {The Journal of Chemical Physics}\ }\textbf {\bibinfo {volume} {92}},\ \bibinfo {pages} {4724} (\bibinfo {year} {1990})}\BibitemShut {NoStop}%
\bibitem [{\citenamefont {Frye}\ and\ \citenamefont {Hutson}(2021)}]{Frye:triatomic-complexes:2021}%
  \BibitemOpen
  \bibfield  {author} {\bibinfo {author} {\bibfnamefont {M.~D.}\ \bibnamefont {Frye}}\ and\ \bibinfo {author} {\bibfnamefont {J.~M.}\ \bibnamefont {Hutson}},\ }\bibfield  {title} {\bibinfo {title} {{Complexes Formed in Collisions Between Ultracold Alkali-metal Diatomic Molecules and Atoms}},\ }\href {https://doi.org/10.1088/1367-2630/ac3ff8} {\bibfield  {journal} {\bibinfo  {journal} {New J. Phys.}\ }\textbf {\bibinfo {volume} {23}},\ \bibinfo {pages} {125008} (\bibinfo {year} {2021})}\BibitemShut {NoStop}%
\bibitem [{\citenamefont {Gao}(2000)}]{Gao:2000}%
  \BibitemOpen
  \bibfield  {author} {\bibinfo {author} {\bibfnamefont {B.}~\bibnamefont {Gao}},\ }\bibfield  {title} {\bibinfo {title} {{Zero-energy Bound or Quasibound States and Their Implications for Diatomic Systems with an Asymptotic van der {W}aals interaction}},\ }\href {https://doi.org/10.1103/PhysRevA.62.050702} {\bibfield  {journal} {\bibinfo  {journal} {Phys. Rev. A}\ }\textbf {\bibinfo {volume} {62}},\ \bibinfo {pages} {050702(R)} (\bibinfo {year} {2000})}\BibitemShut {NoStop}%
\bibitem [{\citenamefont {Chin}\ \emph {et~al.}(2010)\citenamefont {Chin}, \citenamefont {Grimm}, \citenamefont {Julienne},\ and\ \citenamefont {Tiesinga}}]{chengchin}%
  \BibitemOpen
  \bibfield  {author} {\bibinfo {author} {\bibfnamefont {C.}~\bibnamefont {Chin}}, \bibinfo {author} {\bibfnamefont {R.}~\bibnamefont {Grimm}}, \bibinfo {author} {\bibfnamefont {P.}~\bibnamefont {Julienne}},\ and\ \bibinfo {author} {\bibfnamefont {E.}~\bibnamefont {Tiesinga}},\ }\bibfield  {title} {\bibinfo {title} {{Feshbach Resonances in Ultracold Gases}},\ }\href {https://doi.org/10.1103/RevModPhys.82.1225} {\bibfield  {journal} {\bibinfo  {journal} {Rev. Mod. Phys.}\ }\textbf {\bibinfo {volume} {82}},\ \bibinfo {pages} {1225} (\bibinfo {year} {2010})}\BibitemShut {NoStop}%
\bibitem [{\citenamefont {Huo}\ and\ \citenamefont {Green}(1996)}]{Huo:1996}%
  \BibitemOpen
  \bibfield  {author} {\bibinfo {author} {\bibfnamefont {W.~M.}\ \bibnamefont {Huo}}\ and\ \bibinfo {author} {\bibfnamefont {S.}~\bibnamefont {Green}},\ }\bibfield  {title} {\bibinfo {title} {{Quantum Calculations for Rotational Energy Transfer in Nitrogen Molecule Collisions}},\ }\href@noop {} {\bibfield  {journal} {\bibinfo  {journal} {J. Comp. Phys.}\ }\textbf {\bibinfo {volume} {104}},\ \bibinfo {pages} {7572} (\bibinfo {year} {1996})}\BibitemShut {NoStop}%
\end{thebibliography}%

\end{document}